\documentclass[%
reprint,
superscriptaddress,
 amsmath,amssymb,
 aps,longbibliography
]{revtex4-1}
\usepackage{graphicx}
\usepackage{dcolumn}
\usepackage{bm}
\usepackage{color}
\usepackage{blindtext}
\usepackage{hyperref}
\hypersetup{
    bookmarks=true,         
    unicode=false,          
    pdftoolbar=true,        
    pdfmenubar=true,        
    pdffitwindow=false,     
    pdftitle={Certificate},    
    pdfauthor={Dr. F.Sgrignuoli},     
    pdfsubject={TEQIP certificates},   
    pdfcreator={Dr. F.Sgrignuoli},   
    pdfproducer={},  
    pdfkeywords={Certificates,} {TEQIP} {Participation}, 
    pdfnewwindow=true,      
    colorlinks=false,       
    linkcolor=red,          
    citecolor=green,        
    filecolor=magenta,      
    urlcolor=cyan           
}
\usepackage{blindtext}

\begin{document}
\setlength{\parskip}{0.09cm}
\title{Observation of multifractality of light in prime number arrays}
\author{Fabrizio Sgrignuoli}
\affiliation{Department of Electrical \& Computer Engineering and Photonics Center, Boston University, 8 Saint Mary's St., Boston, Massachusetts, 02215, USA.}
\author{Sean Gorsky}
\affiliation{Department of Electrical \& Computer Engineering and Photonics Center, Boston University, 8 Saint Mary's St., Boston, Massachusetts, 02215, USA.}
\author{Wesley A. Britton}
\affiliation{Division of Materials Science \& Engineering, Boston University, 15 Saint Mary's St. Brookline, Massachusetts, 02446, USA}
\author{Ran Zhang}
\affiliation{Division of Materials Science \& Engineering, Boston University, 15 Saint Mary's St. Brookline, Massachusetts, 02446, USA}
\author{Francesco Riboli}
\affiliation{Instituto Nazionale di Ottica, CNR, Via Nello Carrara 1, 50019 Sesto Fiorentino Italy.}
\affiliation{European Laboratory for Nonlinear Spectroscopy, via Nello Carrara 1, 50019 Sesto Fiorentino Italy}
\author{Luca Dal Negro}
\email{dalnegro@bu.edu}
\affiliation{Department of Electrical \& Computer Engineering and Photonics Center, Boston University, 8 Saint Mary's St., Boston, Massachusetts, 02215, USA.}
\affiliation{Division of Materials Science \& Engineering, Boston University, 15 Saint Mary's St. Brookline, Massachusetts, 02446, USA}
\affiliation{Department of Physics, Boston University, 590  Commonwealth Ave., Boston, Massachusetts, 02215, USA}

\maketitle

\textbf{Many natural patterns and shapes, such as meandering coastlines, clouds, or turbulent flows, exhibit a characteristic complexity mathematically described by fractal geometry \cite{Mandelbrot}. In recent years, the engineering of self-similar structures in photonics and nano-optics technology \cite{Berry,Soljacic,Ilday,Cao,Bandres,Cesaro} enabled the manipulation of light states beyond periodic \cite{Joannopoulos} or disordered systems \cite{Wiersma,Lagendijk}, adding novel functionalities to complex optical media \cite{Vardeny,DalNegro_Deter,Torquato,Segev_rev} with applications to nano-devices and metamaterials \cite{Shalaev,Zayats,Zheludev}. Here, we extend the reach of ``fractal photonics" by experimentally demonstrating  multifractality of light in engineered arrays of dielectric nanoparticles. Our findings stimulate fundamental questions on the nature of transport and localization of wave excitations with multi-scale fluctuations beyond what is possible in traditional fractal systems \cite{Berry,Soljacic,Ilday,Cao,Bandres,Cesaro}. Moreover, our approach establishes structure-property relationships that can readily be transferred to planar semiconductor electronics \cite{Khajetoorians} and to artificial atomic lattices \cite{Bloch}, enabling the exploration of novel quantum phases and many-body effects that emerge directly from fundamental structures of algebraic number theory.}

Critical phenomena in disordered quantum systems have been the subject of intense theoretical and experimental research leading to the discovery of multifractality (MF)--intertwined sets of fractals \cite{Stanley,Chhabra,Parisi}--in electronic wave functions at the metal-insulator Anderson transition for conductors \cite{Richardella,Evers}, superconductors \cite{Zhao}, as well as atomic matter waves \cite{Chab}. MF of classical waves has also been observed in the propagation of surface acoustic waves on quasi-periodically corrugated structures \cite{Desideri} and in ultrasound waves through random scattering media close to the Anderson localization threshold \cite{Faez}. Considering the fundamental analogy between the behavior of electronic and optical waves \cite{Lagendijk}, the question naturally arises on the possibility to experimentally observe and characterize multifractal optical resonances in the visible spectrum using engineered photonic media. Besides the fundamental interest, multifractal optical waves offer a novel mechanism to transport and resonantly localize photons at multiple length scales over extended surfaces, enhancing light-matter interactions across broad frequency spectra. These are important attributes for the development of more efficient light sources, optical sensors, and nonlinear optical components \cite{MaciaBook,DalNegro_Deter}. However, to the best of our knowledge, the direct experimental observation of multifractal optical resonances is still missing.  

\begin{figure*}[t!]
\centering
\includegraphics[width=\textwidth]{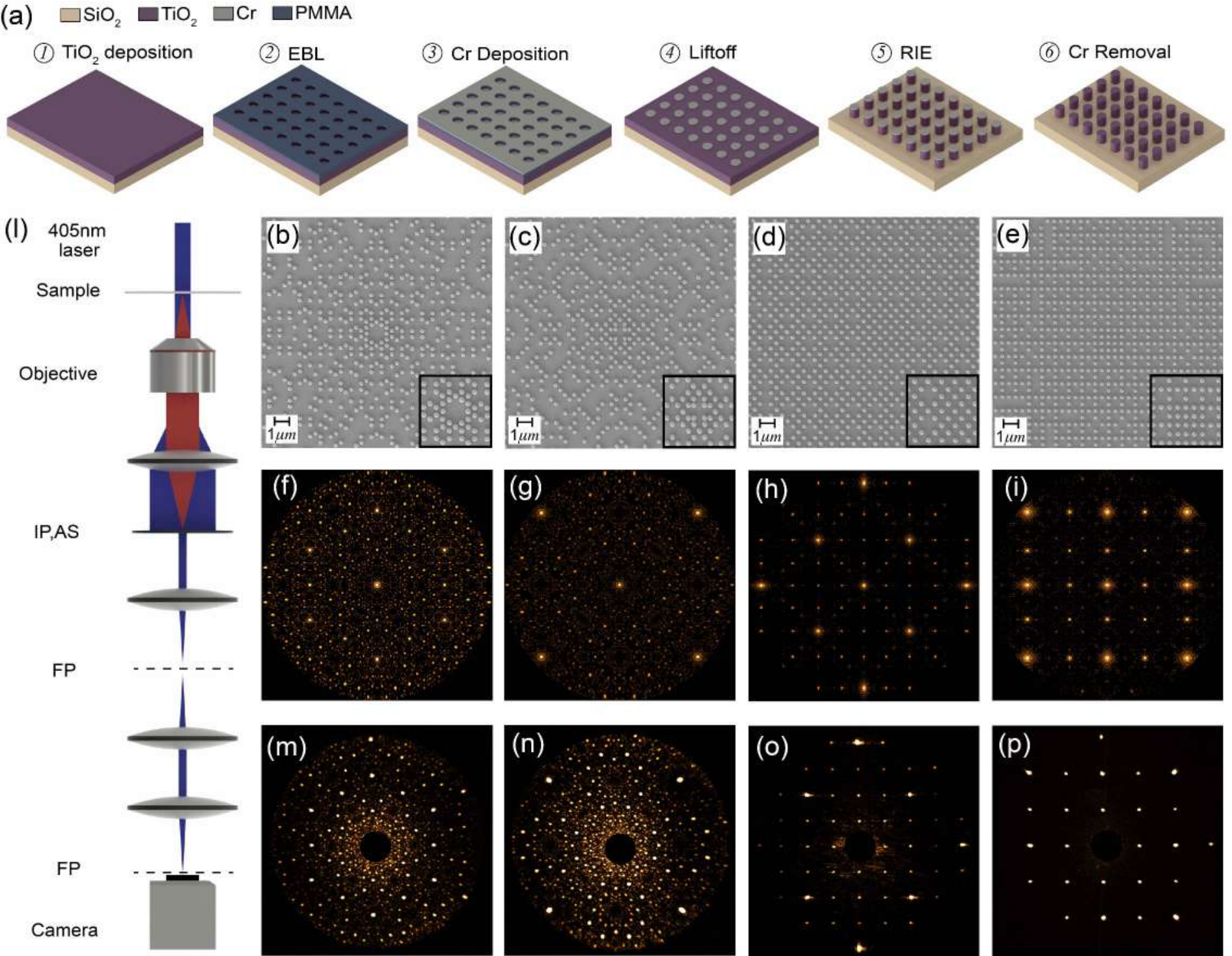}
\caption{Experimental realization of sub-wavelength prime arrays and their diffractive properties. (a) Process flow used to fabricate the TiO$_2$ nanocylinders on a SiO$_2$ substrate. (b-e) SEM images of fabricated samples: panel (b-e) refer to Eisenstein, Gaussian, Hurwitz, and Lifschitz prime configurations, respectively. Insets: enlarged view of the central features for each pattern type. Nanocylinders have a 210\,nm mean diameter, 250\,nm height, and average inter-particle separation of 450\,nm (see also Fig.\,S2). (l) Optical setup for measuring far-field diffraction patterns of laser illuminated sample.  IP=Image Plane, AS=Aperture Stop, FP=Fourier Plane. Calculated (f-i) and measured (m-p) k-space intensity profiles corresponding to the prime arrays reported in panels (b-e), respectively.}
\label{Fig1}
\end{figure*}

In this paper, we demonstrate and systematically characterize multifractality in the optical resonances of aperiodic arrays of nanoparticles that manifest the distinctive aperiodic order intrinsic to fundamental structures of algebraic number theory \cite{Lang,Dekker,Conway}. The devices consist of TiO$_2$ nanopillars deposited atop a transparent SiO$_2$ substrate and arranged according to the prime elements of the Eisenstein and Gaussian integers \cite{Dekker}, as well as two-dimensional cross sections of the irreducible elements of the Hurwitz and Lifschitz quaternions \cite{Conway}. These arrays have recently been shown to support spatially complex resonances with critical behavior akin to localized modes near the Anderson transition in random systems \cite{Faez,Evers}. Fig.\,\ref{Fig1}\,(a) outlines the process flow utilized for the fabrication of the arrays (details in Methods). Scanning electron microscope (SEM) images of the fabricated devices are reported in panels (b-e), while their geometrical properties, illustrated in Figs.\,S1 and S2, are discussed in the Supplementary Information. 
The multifractality in the optical response of these novel photonic structures is demonstrated experimentally by combining diffraction spectra and scattering microscopy across the visible spectral range. 

When laser light illuminates the arrays at normal incidence, sub-wavelength pillars behave as dipolar scattering elements and the resulting far-field diffraction pattern is proportional to the structure factor defined by:
\begin{equation}
S(\bm{k})=\frac{1}{N}\Biggl|\sum_{j=1}^N e^{-i\bm{k}\cdot\bm{r}_j}\Biggr|^2
\end{equation}
where $\bm{k}$ is the in-plane component of the wavevector and $\bm{r}_j$ are the vector positions of the $N$ nanoparticles in the array. This is demonstrated in Fig.\,\ref{Fig1} where we compare the calculated structure factors, shown in panels (f-i), with the measured diffraction patterns, shown in panels (m-p). The optical setup used to measure the diffraction is schematically illustrated in Fig.\,\ref{Fig1}\,(l) and further discussed in the Methods section. The experimental diffraction patterns match very well with the calculated ones and display sharp diffraction peaks embedded in a weaker diffuse background, particularly noticeable in the case of Gaussian and Eisenstein primes. The coexistence of sharp diffraction peaks with a continuous background is characteristic of singular continuous spectra described by singular functions that oscillate at every length scale \cite{Wang_Prime,Baake}. Systems with singular continuous spectra do not occur in nature and are generally associated to multifractal structures. See also Fig.\,S3 and the related discussion. The strength of the continuous spectral component weakens progressively from Eisenstein and Gaussian primes to Hurwitz and Lifschitz structures, which display a more regular geometry.
\begin{figure*}[t!]
\centering
\includegraphics[width=\textwidth]{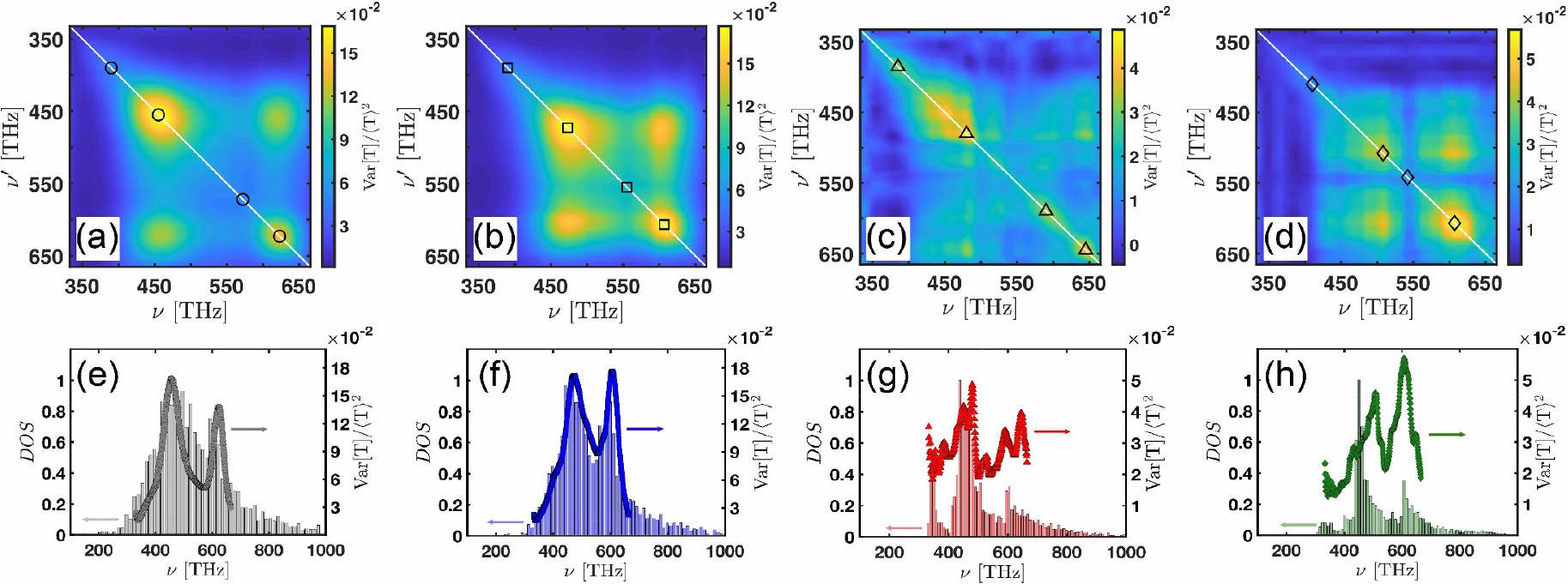}
\caption{Frequency-frequency correlation analysis. (a-d) Frequency-frequency correlation matrix $C(\nu,\nu^\prime)$ of Eisenstein, Gaussian, Hurwitz, and Lifschitz arrays, respectively. The different markers identify the frequency location of representative scattering resonances reported in Fig.\,\ref{Fig3}. (e-f) Normalized variance as compared to the normalized density of states.}
\label{Fig2}
\end{figure*}

Multifractality is demonstrated by analyzing the local scale-invariance of the measured scattering resonances over a large range of scales. In contrast to traditional random media, where MF is achieved only close to the Anderson transition \cite{Richardella,Evers,Faez}, in complex prime arrays MF manifests over a much broader spectral range \cite{Wang_Prime}. 
To establish this fact experimentally, we collected multispectral images of the scattering resonances across the visible spectrum and analyzed their statistical correlations using the frequency-frequency correlation matrix \cite{RiboliPRL}:
\begin{equation}\label{Correlation}
C(\nu,\nu^\prime)=\frac{\langle\delta I(\bm{r},\nu)\delta I(\bm{r},\nu^\prime)\rangle}{\langle I(\bm{r},\nu)\rangle\langle I(\bm{r},\nu^\prime)\rangle}
\end{equation}
where $\bm{r}=(x,y)$ indicates the in-plane coordinates and $\delta I(\bm{r},\nu)=I(\bm{r},\nu)-\langle I(\bm{r},\nu)\rangle$ describes the fluctuations of the intensity with respect to the average value (see Supplementary Information). The degree of spatial similarity between different scattering resonances is directly quantified by the off-diagonal elements of matrix (\ref{Correlation}).

The results of the correlation analysis separate the set of collected scattering resonances into two spectral classes. The first class includes the Eisenstein and Gaussian arrays and is characterized by large fluctuations in the correlation matrix concentrated around two distinct spectral regions, as shown in Fig.\,\ref{Fig2}\,(a-b). In the second class, which includes the Hurwitz and Liftschitz arrays, the $C(\nu,\nu^\prime)$ matrices are more structured and exhibit multi-scale fluctuations spreading over the entire frequency spectrum, as shown in Fig.\,\ref{Fig2}\,(c-d). This characteristic behavior indicates that the spatial intensity distributions of the measured scattered radiation rapidly fluctuate with frequency due to the excitation of scattering resonances in the systems (see Fig.\,S4). This is confirmed by comparing the behavior of the frequency-frequency correlation with the spectral behavior of the computed optical density of states (DOS) of the systems, which we  obtained using the Green's matrix spectral method (see Fig.\,S5 and detailed derivation in the Supplementary Information). Fig.\,\ref{Fig2}\,(e-h) presents the comparison between the DOS and the normalized variance $C(\nu,\nu)$ (see Methods). The Eisenstein and Gaussian prime arrays feature a DOS with two main peaks demonstrating that their scattering resonances are indeed located within the two spectral regions identified using the correlation analysis. On the other hand, the behavior of $C(\nu,\nu)$ and of the DOS for the Hurwitz and Lifschitz configurations features multiple spectral regions of strong intensity fluctuations. 
\begin{figure*}[t!]
\centering
\includegraphics[width=\textwidth]{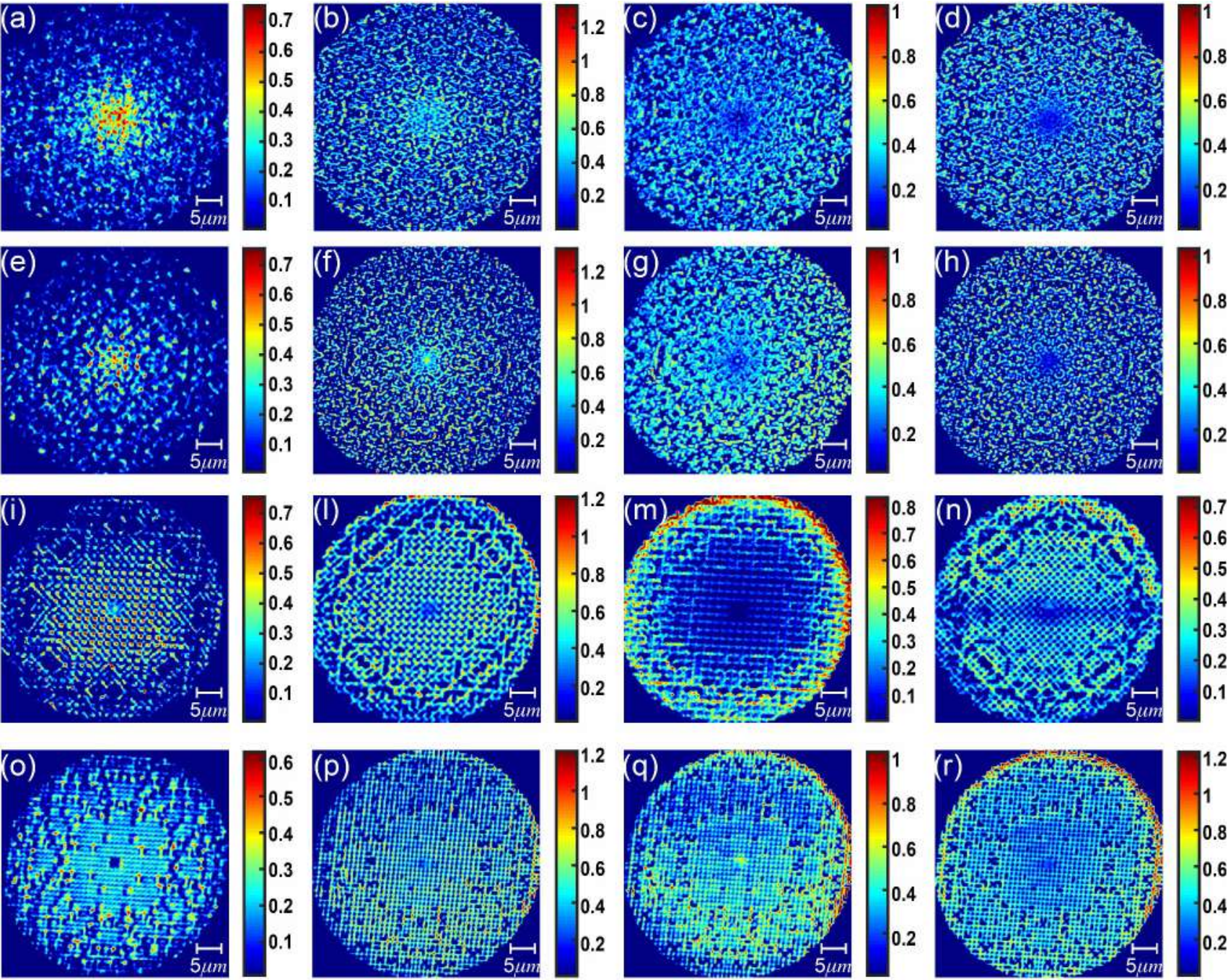}
\caption{Direct observation of the scattering resonances of prime arrays. Representative multispectral dark-field images of the scattering resonances of Eisenstein, (a-d), Gaussian (e-h), Hurwitz (i-n), and Lifschitz (o-r) prime arrays. These quasi-modes are the scattering resonances that interact mostly with the structures and correspond to the spectral positions identified by the markers in Fig.\,\ref{Fig4}\,(a-d). These scattering resonances are normalized with respect to the transmission signal of a reference Ag-mirror.}
\label{Fig3}
\end{figure*}

We now investigate the spatial distributions of the measured scattering resonances within the two identified spectral classes. Fig.\,\ref{Fig3} shows representative dark-field images of the scattering resonances of Eisenstein (a-d), Gaussian (e-h), Hurwitz (i-n), and Lifschitz (o-r) arrays. Their spectral locations are labelled by the symbols in Fig.\,\ref{Fig2}\,(a-d) where different markers capture the main features identified using the correlation analysis. Specifically, the scattering resonances of the Eisenstein and Gaussian prime arrays, which belong to the first spectral class discussed above, display a clear transition from localization in the center of the arrays to a more extended nature in the plane of the arrays (see also Fig.\,S6). This scenario is far richer for the Hurwitz and Lifschitz configurations. In particular, the spatial distributions of their scattering resonances were found to be: (i) weakly localized around the central region of the arrays, (ii) localized at the edges of the arrays, and (iii) spatially extended over the whole arrays (more details provided in Figs.\,S6 and S7). The complex spatial distributions of the scattering resonances shown in Fig.\,\ref{Fig3} exhibit oscillations at multiple length scales, which are characteristic of critical modes in aperiodic systems with fractal and multifractal energy spectra. However, in order to quantitatively describe this behavior, we have analyzed the local scaling of the measured spatial intensity distributions using rigorous multifractal analysis \cite{Chhabra}. In particular, we employ the box-counting method to characterize the size-scaling of the moments of the light intensity distribution. This is achieved by dividing the system into small boxes of varying size $l$. We then determine the minimum number of boxes $N(l)$ needed to cover the system for each size $l$ and evaluate the fractal dimension $D_f$ using the power-law scaling $N(l)\sim l^{-D_f}$. Traditional (homogeneous) fractal structures are characterized by a global scale-invariance symmetry described by a single fractal dimension $D_f$. On the other hand, heterogeneous fractals or multifractals are characterized by a continuous distribution $f(\alpha)$ of local scaling exponents such that $N(l)\sim l^{-f(\alpha)}$ \cite{Chhabra}. The so-called singularity spectrum $f(\alpha)$ generalizes the fractal description of complex systems in terms of intertwined sets of traditional fractal objects. Different approaches are used to extract $f(\alpha)$ from the local scaling analysis. We have obtained the multifractal spectra directly from the dark-field measurements by employing the method introduced in ref. \cite{Chhabra}. Details on the implementation are discussed in the Supplementary Information and in Fig.\,S7-S11. 
\begin{figure*}[t!]
\centering
\includegraphics[width=\textwidth]{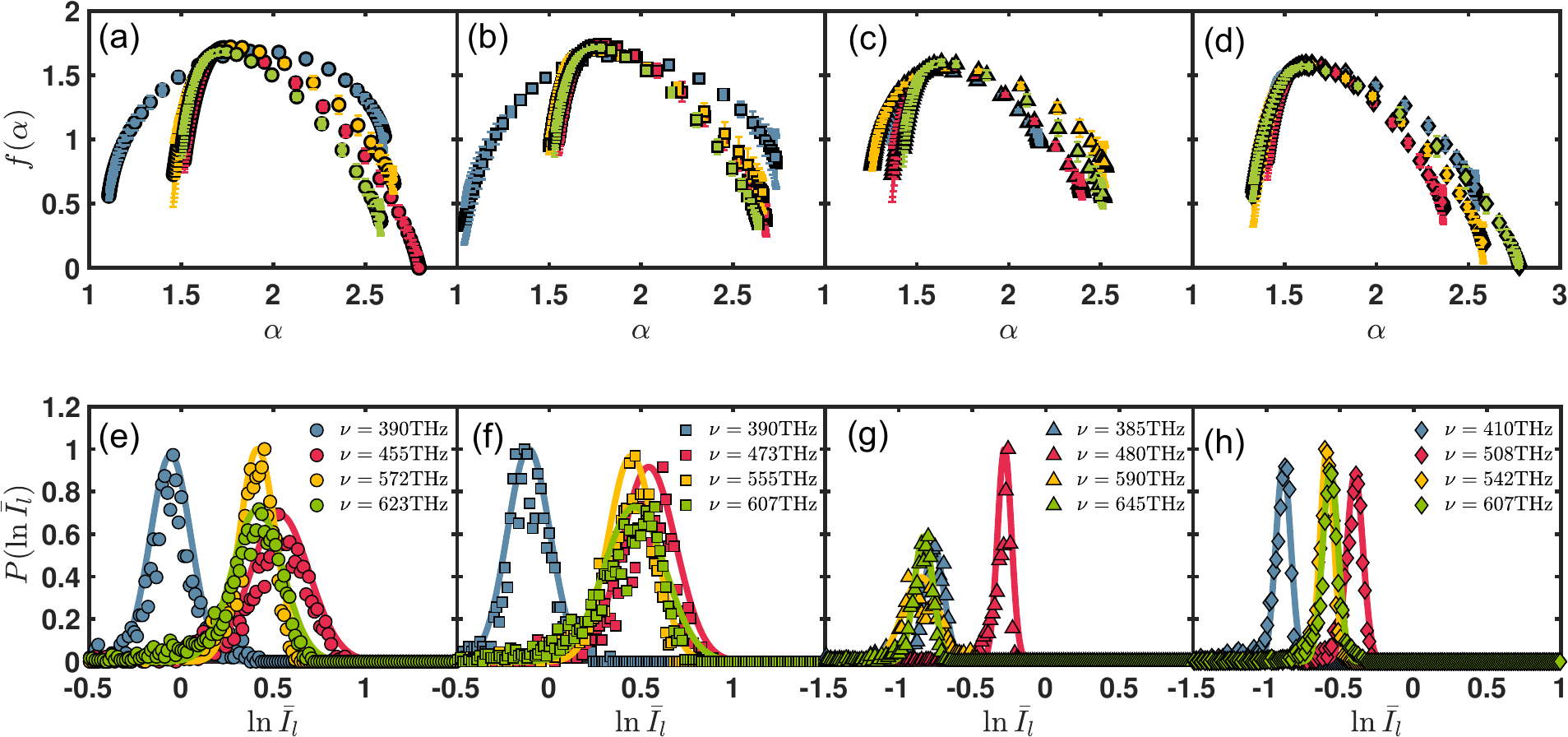}
\caption{Multifractal nature of scattering resonances of prime arrays. Multifractal singularity spectra (a-d) and probability density functions of the box-integrated intensity $P(\ln I_l)$ (e-h) of representative scattering resonances of Eisenstein (a,e), Gaussian (b,f), Hurwitz (c,g), and Lifschitz (d,h) prime arrays, respectively. Their resonance spectral positions are identified by the different markers of Fig.\,\ref{Fig2}\,(a-d). The continuous lines in panels (e-h) show the best fits obtained using log-normal function model.}
\label{Fig4}
\end{figure*}

Fig.\,\ref{Fig4}\,(a-d) demonstrate the multifractal nature of the measured scattering resonances of the prime arrays. All the singularity spectra $f(\alpha)$ exhibit a downward concavity with a large width $\Delta\alpha$, which is the hallmark of multifractality \cite{Chhabra}. The behavior of the width $\Delta\alpha$ as a function of frequency reflects the previously identified classification in terms of the correlation matrix (see Fig.\,S9). In particular, the arrays with more significant intensity fluctuations, i.e. Eisenstein and Gaussian arrays, also display the broadest singularity spectra. Our findings establish a direct connection between the multifractal properties of the geometrical supports of the arrays (discussed in the Supplementary Information), the MF of the corresponding scattering resonances, and the singular continuous nature of diffraction patterns. To further characterize the MF of the scattering resonances we also computed, based on the experimental data, the mass exponent $\tau(q)$ and the generalized dimension $D(q)$, which provide alternative descriptions of multifractals. Multifractals are unambiguously characterized by a nonlinear $\tau(q)$ function and a smooth $D(q)$ function \cite{Stanley}. This is reported in Fig.\,S8 and discussed in the Supplementary Information.

A direct consequence of multifractality is the non-Gaussian nature of the probability density function (PDF) of the light intensity distribution near, or at criticality \cite{Richardella,Faez}. We have evaluated the PDF of the scattered radiation from the histogram of the logarithm of the box-integrated intensities, i.e. $P(\log[I_{l_i}])$. We show in Fig.\,\ref{Fig4}\,(i-l) the histograms produced by a box size of approximately $0.8\times\lambda$ where $\lambda$ is the wavelength in each case. We note that, although the intensity distributions of the scattering resonances in Fig.\,\ref{Fig3} show different degrees of spatial variations, all the PDFs are very-well reproduced considering log-normal distributions (continuous lines in Fig.\,\ref{Fig4}\,(e-h)). Moreover, we have verified that these findings do not depend on the box-counting size $l$, as shown in Fig.\,S10. Furthermore, the multifractal spectrum $f(\alpha)$ can be rigorously obtained from the PDF of the intensity within the parabolic approximation:
\begin{equation}\label{PA}
f(\alpha)=D_f-(\alpha-\alpha_0)^2/4(\alpha_0-d)
\end{equation}
where $D_f=f(\alpha_0)$ is the fractal dimension of the geometrical support \cite{Faez}. Equation\,(\ref{PA}) characterizes $f(\alpha)$ by the single parameter $\alpha_0$ associated to the mean $\mu$ and variance $\sigma$ of the PDF via the formula $\alpha_0=2\mu d/(2\mu+\sigma^2)$. Deviations from the parabolic spectrum are associated to the strong MF of the scattering resonances of prime arrays across a broad range of frequencies (see Supplementary Information). The observed strong MF in the scattering resonances is uniquely associated to the multi-scale geometrical structure of prime arrays. In contrast, only weak MF was previously reported in uniform random scattering media at their metal-insulator Anderson transitions \cite{Richardella,Faez}.
 
In conclusion, we have provided the first experimental observation of multifractality of classical waves in the optical regime by studying the resonances of photonic structures designed to reflect the intrinsic aperiodic order of complex primes and the irreducible elements of quaternion rings. Beyond its fundamental interest with respect to the discovery and characterization of multifractality in number theory \cite{Oliver,Wolf1}, our findings establish a novel mechanism to transport and resonantly localize photons at multiple length scales and to enhance light-matter interactions with applications to active photonic devices and novel broadband nonlinear components. Moreover, the concepts developed in this work can be naturally transferred to quantum waves \cite{Bloch,Viebahn} and stimulate the development of novel quantum phases of matter and many-body phenomena that emerge from fundamental structures of algebraic number theory.
\section*{Methods}
\subsection*{Sample fabrication}
TiO$_2$ thin films were grown by reactive DC magnetron sputtering (MSP) on quartz (SiO$_2$) substrates with a Denton Discovery D8 Sputtering System. A 3 inch diameter Ti target (Kurt J. Lesker, 99.998$\%$) was used. The deposition was performed under a 200 Watt power at a 3.0 mTorr deposition pressure with a 1:3 Sccm flow rate ratio of O$_2$ to Ar gases. These growth conditions resulted in a deposition rate of about 1\,nm/min. Chamber base pressure was kept below 5$\times$10$^-7$ Torr, the target-substrate distance was fixed at 10cm, and substrates were rotated at a speed of 5\,rpm. Substrates were solvent washed and plasma cleaned in O$_2$ prior to deposition.

The optical constants of the  films were determined using a variable angle spectroscopic ellipsometer (V-Vase, J.A. Woollam) in the wavelength range of 300\,nm to 2000\,nm. The measured data were fitted in good agreement with the Cauchy model, an empirical relationship between refractive index and wavelength. The film thickness, verified with ellipsometry and scanning electron microscopy (SEM), was targeted at 250\,nm.

Nanoparticle fabrication was performed with electron beam lithography. A PMMA resist layer of about 100\,nm thickness was spun and baked before sputtering a thin conducting layer ($\sim$6\,nm) of Au to compensate for the electrically insulating SiO$_2$ substrate. The resist was exposed at 30 keV by an SEM (Zeiss Supra 40) integrated with Nanometer Pattern Generation System direct write software. The sample was then developed in IPA:MIBK (3:1) solution for 70sec followed by a rinsing with IPA for 20sec. Next, a 20\,nm thick layer of Cr was deposited on top of the developed resist with electron beam evaporation (CHA Industries Solution System). After this deposition, unwanted Cr was removed with a three minutes lift-off in acetone, and the nanoparticle patterns were transferred from the Cr mask to the TiO$_2$ thin film by reactive ion etching (RIE, Plasma-Therm, model 790) using Ar and SF$_6$ gases. Finally, the Cr mask was removed by wet etching in Transene 1020 \cite{Zheng}.
\subsection*{Optical characterization}
We measure the far-field diffraction pattern of laser light passing through the sample using the setup depicted in Fig.\,\ref{Fig2}\,(l). A 405\,nm laser is focused onto the patterned area to overfill the pattern and to ensure uniformity in intensity across it. The forward scattered light is collected by a high numerical aperture objective (NA=0.9 Olympus MPlanFL N) which collects light scattered up to $64^\circ$ from the normal direction. A 4-F optical system, immediately behind the objective, creates an intermediate image plane and intermediate Fourier plane. An iris located at the intermediate image plane was used to restrict the light collection area only to the patterned region. A second 4-F optical system then re-images the intermediate Fourier plane onto the CCD (MediaCybernetics) with the appropriate magnification. Finally, digital filtering was employed to remove the strong D.C. component of the diffraction spectra to produce clear images. In order to measure the spatial distribution of the scattering resonances of prime arrays, we used a line-scan hyperspectral imaging system (XploRA Plus by Horiba Scientific). The field of view is illuminated by light from a lamp through the microscope objective.  A line-like region of the field of view is projected into the spectrograph entrance slit. The slit image is then spectrally dispersed by means of a grating onto the camera image sensor. This allows the collection of all the spectra corresponding to a single line in the field of view. Scanning the sample with a stage then produces the hyperspectral image of the entire field of view. In our system, scanning 696 lines results in square-size hyperspectral image. The resolution of the CCD (Cooke/PCO Pixelfly PCI Camera) was set to be 430$\times$470, where 430 and 470 corresponds to a spatial and spectral resolution of 30\,nm and 0.25\,THz, respectively. The exposure time was set to be 1000\,ms with a time interval of 1\,min between different scanning lines. Finally, the dark-field data were normalized with respect to a reference signal of a Ag-mirror. 
 \subsection*{Correlation analysis}
Each element $C(\nu_i,\nu_j)$ of the frequency-frequency correlation matrices of Fig.\,\ref{Fig4} (a-d) is the result of an average over $9\times10^4$ correlated values. The normalization of the covariance $\langle\delta I(\bm{r},\nu)\delta I(\bm{r},\nu^\prime)\rangle$ with respect to the product of the average values $\langle I(\bm{r},\nu) I(\bm{r},\nu^\prime)\rangle$ minimizes the intrinsic spectral effects related to the illumination source of the experimental apparatus (see equation\,(\ref{Correlation})). Each element of $C(\nu,\nu^\prime)$ is a combination of spatial intrinsic fluctuations of the system's parameters and extrinsic effects related to the illumination-collection efficiency and to the point spread function of the experimental apparatus. The spectral dependence of the extrinsic effects can be mitigated by the proper normalization of the correlations matrix \cite{RiboliPRL}. 
The white diagonal elements of Fig.\,\ref{Fig2} (a-d) as well as the continuous lines in Fig.\,\ref{Fig2} (e-h) are the normalized variance of the scattering resonances of prime arrays and are calculated by the equation:
\begin{equation}\label{Varsignal}
C(\nu,\nu)=\frac{\langle\delta I(\bm{r},\nu)^2\rangle}{\langle I(\bm{r},\nu)\rangle^2}=\frac{\sigma^{2}(\nu)}{\mu^{2}(\nu)}
\end{equation}
where $\mu(\nu)$ is the average value and $\sigma(\nu)$ is the standard deviation. 

\subsection*{Multifractal analysis}
The multifractal analysis of both structural and dynamical properties was performed from the corresponding 600 dpi bitmap image using the direct Chhabra-Jensen algorithm \cite{Chhabra} implemented in the routine FracLac (ver. September 2015) developed for the NIH distributed Image-J software package \cite{Schneider}. 
This method is a useful tool to determine the scaling properties of a certain image, but has to be handled with care. First of all, the size of the boxes must satisfy some requirements. In the FracLac routine, the largest box should be larger than 50\% but not exceed the entire image, while the smallest box is chosen to be the point at which the slope starts to deviate from the linear regime in the $\log(N)$ versus $\log(1/r)$ plot \cite{Schneider}. Furthermore, the scattering resonance maps are not binary. Therefore, it is necessary to specify the threshold value above which the pixels are part of the object under analysis. Different calculations were performed for several threshold percentages (between 55\% and 75\%) of the maximum intensity of each scattering resonance of Fig.\,\ref{Fig3}. Another source of error could be the scaling method used during the analysis. For each dark-field image we employed a linear, a rational, and a power scaled series of box sizes \cite{Schneider}. All these aspects do not affect the main results of our paper, as shown by the error bars of Fig.\,\ref{Fig4} (a-d). 
\section{Acknowledgements}
All the authors would like to acknowledgment Dr. R.Wang for his contribution during the first stage of this work. All the authors would like to acknowledgment Dr. A. Ambrosio and Dr. M. Tamaglione for their help during the multispectral dark-field measurements performed at the Center for Nanoscale Systems at Harvard University (Cambridge, MA 02138 USA). F.S acknowledges Dr. F. Scazza for fruitful discussion on the applications of aperiodic lattices on ultracold quantum physics. L.D.N. acknowledges partial support from the Army Research Laboratory under Cooperative Agreement Number W911NF-12-2-0023 for the development of theoretical modeling.
\section*{Author contributions statement}
F.S. performed numerical calculations, data analysis, organized the results, and contributed to the experimental measurements. S.G. performed Fourier spectra experiments, dark field measurements and contributed in numerical simulations. W.A.B. and R. Z. fabricated the samples. F.R. performed the correlation analysis. L.D.N. conceived, lead and supervised the research activities. L.D.N organized the results. L.D.N. and F.S. wrote the manuscript with critical feedback from all the authors. All authors contributed to discussions and manuscript revision.
\section{Supplementary Information}

\section*{Multifractal analysis of prime arrays}
\begin{figure*}[b!]
\centering
\includegraphics[width=\textwidth]{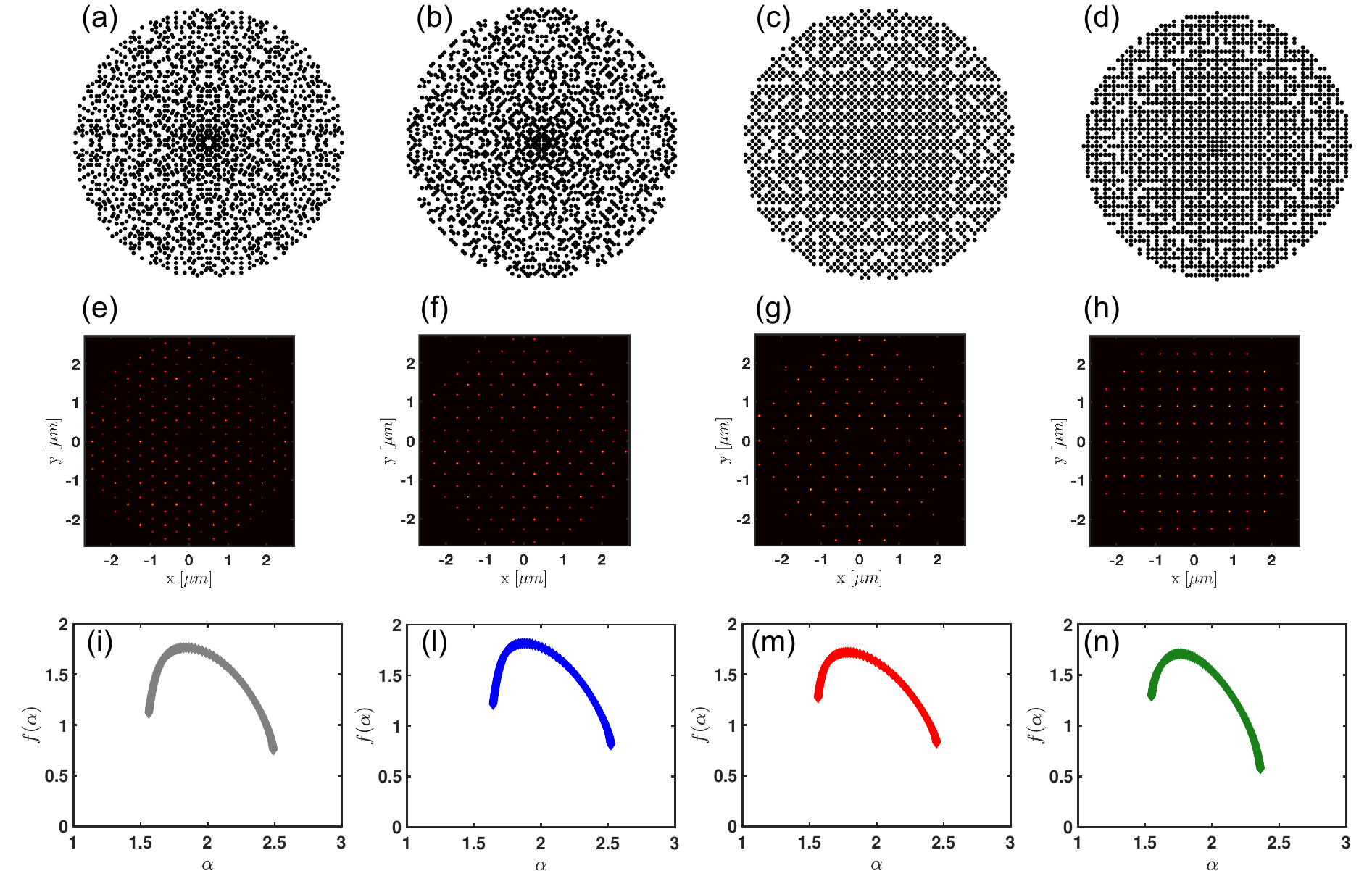}
\caption{Geometrical properties of point patterns based on prime elements distributions. Prime point patterns generated by considering (a) N=2100 Eisenstein primes (EP), (b) N=2016 Gaussian primes (GP), (c) N=1957 Hurwitz primes (HP), and (d) N=2013 Lifschitz primes (LP).(e-h) Vector two-point correlation functions $g_2(r,\theta)$ of the prime arrays, and (i-n) singularity spectra $f(\alpha)$ corresponding to the respective array at the top of each column. The fractal dimensions $D_f$ are equal to 1.76, 1.82, 1.72, and 1.71 for the Eisenstin, Gaussian, Hurwitz, and Lifschitz geometry, respectively }
\label{FigS1}
\end{figure*}

The spatial distributions of the prime arrays considered in the paper are shown in Fig.\,\ref{FigS1}\,(a-d).  The Eisenstein and Gaussian prime arrays, generated from the prime elements in the integer rings of complex quadratic fields, are characterized by a 12-fold and an 8-fold rotational symmetry, respectively. These features are evident from Fig.\,\ref{FigS1}\,(e-f) where we report their vector two-point correlation function $g_2(r,\theta)$, defined by:
\begin{equation}
g_2(r,\theta)dA=\frac{N_{exp}}{\rho}
\end{equation}
where $\rho$ is the averaged point density and $N_{exp}$ refers to the expected number of points contained in the infinitesimal area $dA$=$rdrd\theta$ at position $(r,\theta)$. On the other hand, the Hurtwitz and Lifschitz configurations, obtained from two-dimensional cross section of the irreducible elements of quaternions, can be interpreted as the integer and the half-integer lattice points on a sphere of radius $\sqrt{N(z)}$ in $\mathbb{R}^4$, respectively. These lattice structures are evident in their $g_2(r,\theta)$, as shown in Fig.\,\ref{FigS1}\,(g-h). We refer to refs.\cite{Wang_Prime,Vardi,Rudnick,Prasad,Conway} for more details about the mathematical backgrounds to generate these point pattern distributions and to get more information on their geometrical properties.

To characterize the multifractal properties of prime elements in complex algebraic rings, we have applied the multifractal scaling analysis to the corresponding point patterns. As discussed in the manuscript, multifractal analysis is a statistical description that enables the study of long-term dynamical behavior of an arbitrary signal \cite{Chhabra,Hentschel,Halsey,Meakin,Parisi,Hilborn,Nakayama,Stanley,Mandelbrot,Feder}. The idea of multifractality (MF), first introduced to analyze energy dissipation in turbulent fluids \cite{Parisi}, widened our understanding of complex and intricate distributions observed in various fields of science \cite{Nakayama}. Specifically, multifractal analysis provided a number of significant insights into signal analysis \cite{Muzy}, finance \cite{MandelbrotFinance,Schmitt,Schertzer}, network traffic \cite{Riedi,Taqqu,Ribeiro}, photonics \cite{Albuquerque,Trevino,Macia,Ryu,Sorensen,Sroor,Desideri,DalNegro_Deter}, and critical phenomena \cite{Richardella,Rodriguez,Zhao,Schreiber,Evers,Faez,Stanley,Mandelbrot,Feder,StanleyBook}, to cite a few.

Fig.\,\ref{FigS1}\,(i-n) report the results of this study. The geometrical supports of the prime arrays exhibit clear MF with singularity spectra $f(\alpha)$ of characteristic downward concavity which extends over a compact interval $[\alpha_{min},\alpha_{max}]$. Here, $\alpha_{min}$ (respectively $\alpha_{max}$) corresponds to the weakest (respectively the strongest) singularity. It is important to emphasize that if the geometrical support of an object is fractal (or multifractal), then also their dynamical properties (like their wavefunctions, energy spectra, and local density of states) exhibit fractality (multifractality) \cite{Nakayama, Albuquerque,Trevino,Macia,Ryu,Kempkes}. Interestingly, the opposite is not true. Indeed, electron wavefunctions and energy spectra at the frequency where the Anderson metal-insulator transition occurs have multifractal properties although electrons lie on an Euclidean (non-fractal) support \cite{Nakayama,Schreiber}. 

\begin{figure*}[b!]
\centering
\includegraphics[width=\linewidth]{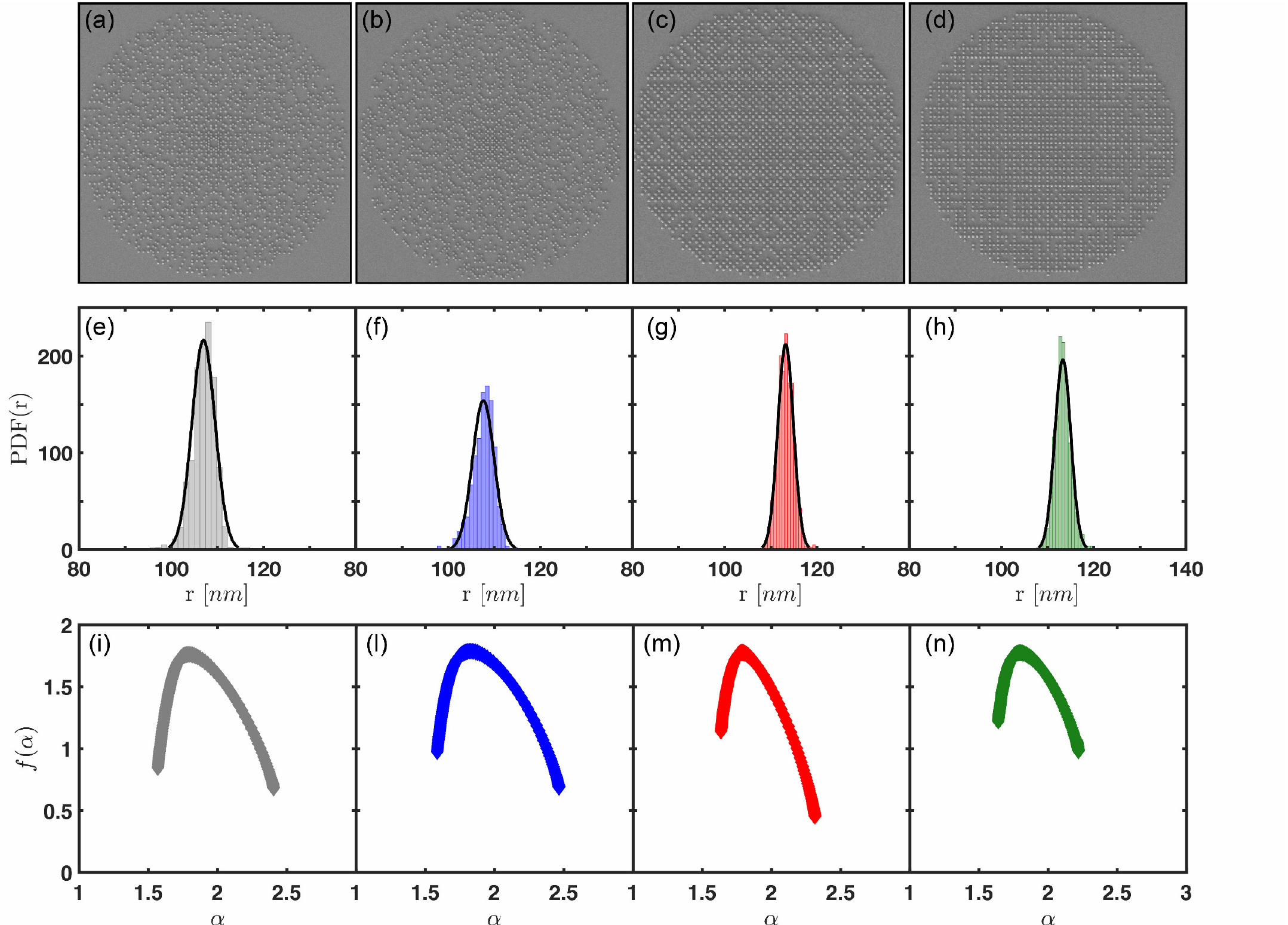}
\caption{Geometrical properties of photonic prime arrays. Scanning Electron Microscope (SEM) of Eisenstein (a), Gaussian (b), Hurwitz (c), and Lifschitz configurations. (e-h) Nanocylinder size distributions of the corresponding prime arrays reported on top of each column evaluated from the SEM images by using the Image-J software package \cite{Schneider}. (i-n) Singularity spectra $f(\alpha)$ of Eisenstin, Gaussian, Hurwitz, and Lifschitz geometry, respectively. The fractal dimensions $D_f$ are equal to 1.76, 1.79, 1.77, and 1.76, respectively.}
\label{FigS2}
\end{figure*}
In order to corroborate MF in the geometrical support of prime arrays, we have also applied multifractal scaling analysis to the SEM images of the fabricated TiO$_2$ nanoparticles deposited atop of a quartz substrate. Fig.\,\ref{FigS2}\,(a-d) show SEM images of the Eisenstein (a), Gaussian (b), Hurwitz (c), and Lifschitz (d) prime distributions. The fabricated nanocylinders have a 210\,nm mean diameter, 250\,nm height, and average inter-particle separation of 450\,nm, as reported in Fig.\,\ref{FigS2}\,(e-d) that display histograms of the nanoparticle size distributions evaluated by using the ImageJ software package \cite{Schneider}. These distributions show the traditional Gaussian profile (black-curves identify the Gaussian fits) associated to the random fluctuations of the electron beam and to the diffusion of the resist during the development. These results demonstrate the homogeneity of the size of the particles in the fabricated devices. Accordingly, the fabricated arrays display the same multifractal features of the point pattern distributions. The corresponding $f(\alpha)$ spectra are reported in Fig.\,\ref{FigS2}\,(i-n). Moreover, the Eisenstein and Gaussian prime arrays show a larger width $\Delta\alpha$, which describes the degree of inhomogeneity of the structures, compared to the Hurwitz and Lifschitz configurations. This is consistent with the more complex nature of their geometrical supports, as also reflected in the features of their $g_2(r,\theta)$.

\begin{figure*}[t!]
\centering
\includegraphics[width=10cm]{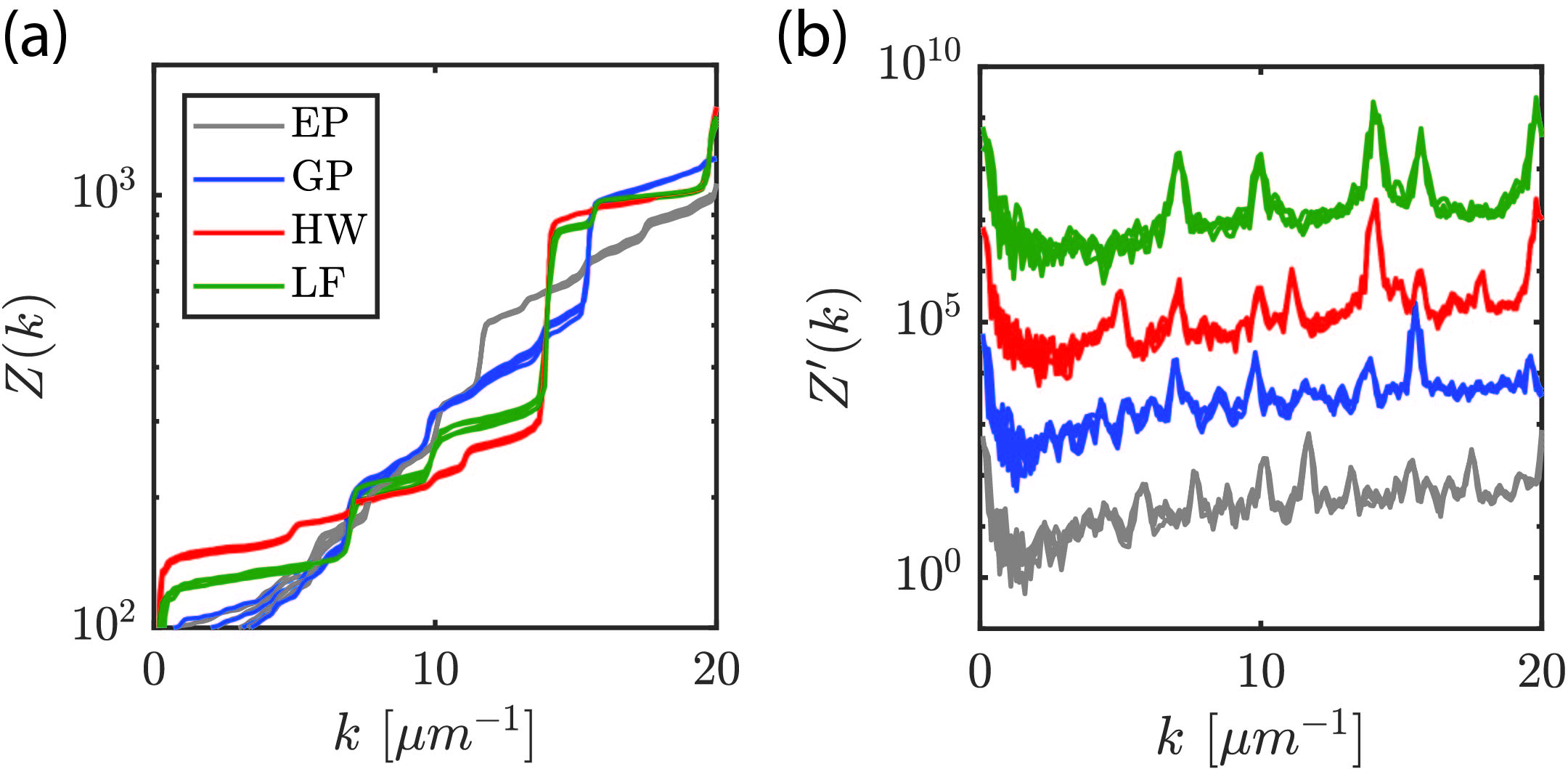}
\caption{Integrated structure factor $Z(k)$ scaling analysis. (a) Z(k) and (b) $Z^\prime (k)$ for the four arrays. In (b) the GP, HW, and LF curves are vertically offest by factors of $10^2$, $10^4$, and $10^6$, respectively.  The non-zero baseline of $Z^\prime (k)$ demonstrates the singular-continuous nature of the structures. In both plots there are four nearly-overlapping curves of each color, corresponding to different particle numbers in the array ranging from 500 to 2000 particles, showing that the result is insensitive to particle number.}
\label{Zkscaling}
\end{figure*}

As discussed in the main text, the MF of a geometrical object is also reflected in the singular continuous nature of its diffraction spectrum \cite{Baake,BaakeTM}. Although this cannot be rigorously proven, the singular continuous nature of the diffraction spectra can be shown by analyzing the cumulative integrated structure factor defined by the relation \cite{Wang_Prime,Baake}:
\begin{equation}
    \label{Zk}
    Z(k)=\int_0^{2\pi}\int_{0}^{k}S(q,\phi)qdqd\phi
\end{equation}
The structure factor $S(\boldsymbol{k})$ of singular-continuous structures consist of bright peaks on a smooth and continuous background \cite{Wang_Prime}. Therefore, $Z(k)$, also named integrated intensity function, will exhibit sharp jumps connected by monotonically increasing intervals. Alternatively, its derivative $Z^\prime (k)$ consists of a non-zero baseline with intermittent spikes. All these features can indeed be observed for the all four prime arrays in Fig.\,\ref{Zkscaling}\,(a,b) when studying $Z(k)$ and $Z^\prime (k)$, respectively. 
An important aspect of this analysis is to make sure that the observed scaling curves are not dependent on the number of particles in the arrays. This is established in Fig.\,\ref{Zkscaling} where we report the scaling of $Z(k)$ and $Z^\prime (k)$ with respect to the size of the different prime arrays. Each curve in panels (a-b) is a superposition of four and nearly overlapping lines corresponding to different number of particles from 500 to 2000 elements (in intervals of 500). The small qualitative changes reported when varying the numbers of particles show that this analysis is very robust with respect to the size of the arrays and is indicative of their singular continuous nature.

\section*{Frequency-frequency correlation analysis.}
Differently from the traditional random media where MF is achieved only close to the Anderson-transition \cite{Richardella,Evers,Faez}, in complex prime arrays MF is exhibited over a much broader spectral range \cite{Wang_Prime}. The frequency-frequency correlation matrices $C(\nu,\nu^\prime)$, reported in Fig.\,2 of the main manuscript, were used to establish a statistical and predictive relationship between the degree of similarity of all the measured dark-field scattering resonances. The degree of spatial similarity is measured by their off-diagonal elements \cite{RiboliPRL}. The main result of this  analysis is the classification of prime arrays into two different classes, which reflects their mathematical structure. The first class comprises the Eisenstein and Gaussian (integers complex prime numbers) arrays, while the second one Hurtwitz and Lifschitz (generated from quaternions numbers) arrays. In the following we will discuss the relation between the $C(\nu,\nu^\prime)$ results and the spatial features of the representative scattering resonances shown in Fig.\,3 of the manuscript.

The first class is characterized by peak intensity fluctuations concentrated in two well-defined spectral regions located at $\nu\approx460$THz and $\nu\approx600$THz, respectively, with very similar absolute values close to $C(\nu,\nu)\approx0.18$. Here, $C(\nu,\nu)$ indicates the normalized variance of the signal (see the main muascript for more details). In the present work, we didn't evaluate the autocorrelation function due to the strong variations of the frequency dispersion of the scattering resonances of prime arrays \cite{RiboliPRL}. The off-diagonal peaks show that these large fluctuating resonances are correlated, indicating that their spatial distributions share common structural features. Moreover, a depletion of the intensity fluctuations is observed around $\nu\approx550$THz in both Eisenstein and Gaussian arrays. All these features can be qualitatively observed also in the spatial distributions of the representative scattering resonances reported in Fig.\,3 of the manuscript. Indeed, the Eisenstein and Gaussian configurations are characterized by a clear transition from modes localized in the center of the arrays to optical resonances more radially spreading in the plane of their geometrical supports. Specifically, Figs.\,3\,(a) and (c) do not share any common features: while the resonance (a) is more concentrated in the center of the arrays, resonance (c) is characterized by a minimum of the intensity in that region. On the other hand, the spatial distributions of panels (b) and (d) are correlated, as also shown by the off-diagonal peaks of the correlation matrix of Fig.\,2\,(a). The same conclusions are obtained in the Gaussian arrays: while the resonances (e) and (g) are not correlated, the spatial distributions (f) and (h) are similar.
\begin{figure*}[t!]
\centering
\includegraphics[width=\linewidth]{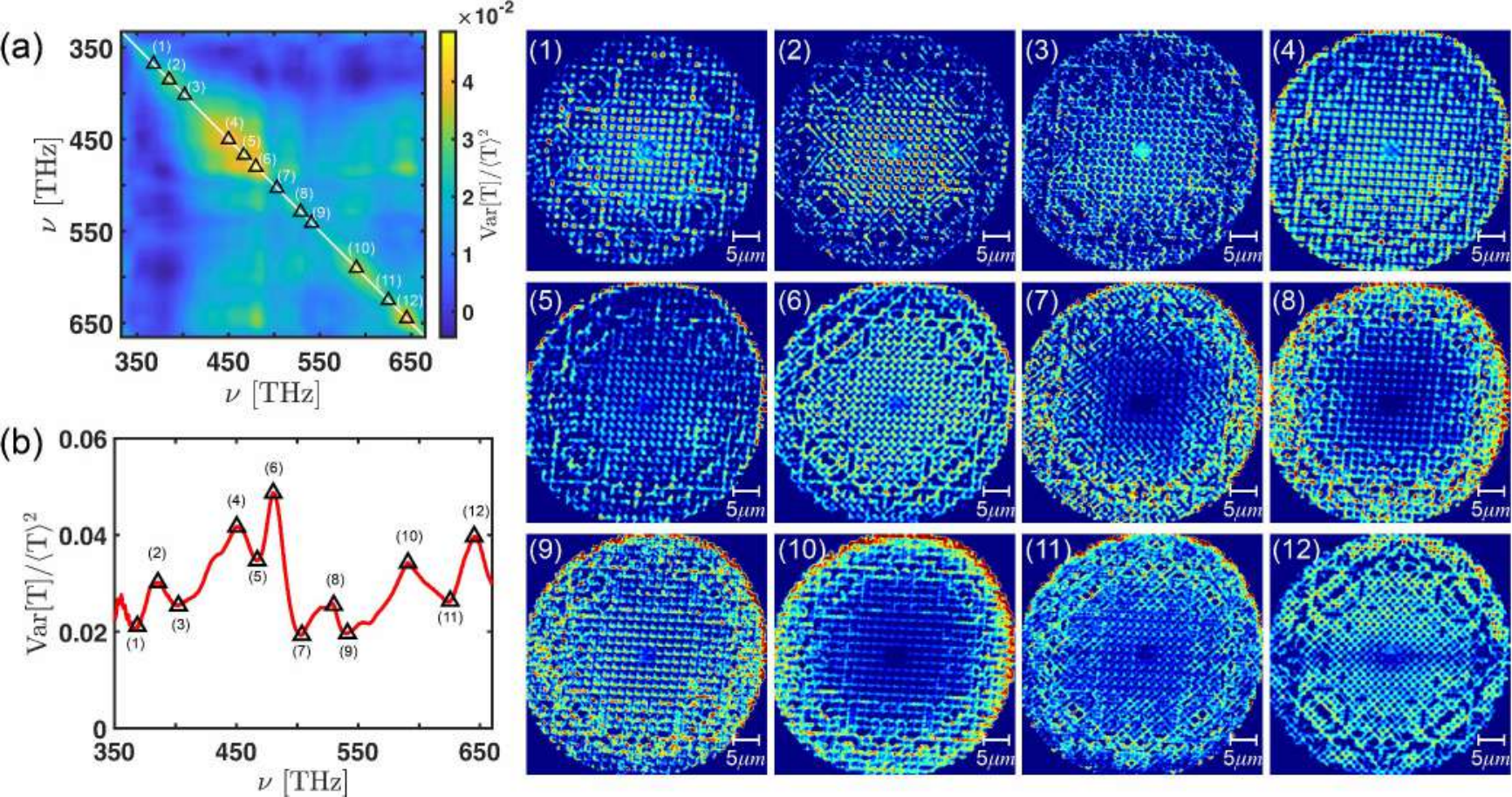}
\caption{Frequency-frequency correlation of Hurwitz prime array. (a) Frequency-frequency correlation matrix $C(\nu,\nu^\prime)$. (b) Normalized variance $C(\nu,\nu)$. The different numbers identify the spectral positions of 12 representative scattering resonances that are also reported in the figure.}
\label{FigS3}
\end{figure*}

The second class displays more variegated and richer correlation matrices. Specifically, their normalized variances are characterized by peaks and dips that spread over the entire measured frequency spectrum. This behavior indicates that their spatial distributions rapidly fluctuate with frequency. In the Hurwitz configuration, for example, some of these fast fluctuating resonances are correlated (or anti-correlated), as highlighted by the off-diagonal elements of the correlation matrix. Indeed, Fig.\,\ref{FigS3} shows that scattering resonances of the Hurwitz array can be classified in four different spectral regions: (i) [350,420] THZ (dark-field images 1-3), (ii) [430,480] THz (dark-field images 4-6), (iii) [480,600] THz (dark-field images 7-11), (iv) larger than 600 THz (dark-field image 12). Specifically, the region (i) is characterized by scattering resonances that are mostly concentered at the center of the array and that are anti-correlated with respect to the all the others, as shown in Fig.\,\ref{FigS3}\,(b). On the other hand, the spectral region (ii) is correlated with region (iv). The spectral region (iii), characterized by a minimum of the fluctuations of the scattered intensity, is populated by scattering resonances spatially localized at the edge of their geometrical support. Interestingly, these optical modes can be the analogues of the recently discovered topological edge states in aperiodic system \cite{Bandres,Wang_OpticsLett}. The same considerations can be applied to the Lifschitz prime array. Though less noticeable, clear features of edge states can be recognized in Fig.\,3\,(q), which depicts a representative scattering resonance spectrally located in a minimum of its normalized variance. Furthermore, the scattering resonances (o) and (q) of Fig.\,3 are not correlated in space, while (p) and (r) are correlated. This complex scenario is well-reproduced in our numerical simulations shown in Fig.\,\ref{FigS5} (see the next section for more details). Therefore, the quasi-modes of this second class display a more complex spatial structure as compared to Eisenstein and Gaussian arrays. In addition, the absolute value of their fluctuations ($C(\nu,\nu)\approx0.05$) are smaller then the ones of the first class consistently with the more uniform nature of Hurwitz and Lifschitz arrays. 
\section*{Electric and magnetic Green's matrix method}
In this section, we discuss an extension of the Green's matrix spectral method \cite{RusekPRA}, named electric and magnetic coupled dipole approximation (EMCDA) \cite{DalNegro_Crystals2019}. This theoretical approach accounts for both the first-order Mie-Lorentz coefficients through the induced electric and magnetic polarizability \cite{Yurkin} and it provides valuable physical insights into the complex behavior on how light interacts with the aperiodic prime arrays. This approximation enables the simulation of the scattering response of large-scale photonic arrays with several thousand interacting particles, which are well beyond the reach of traditional numerical methods such as finite difference time domain (FDTD) or finite element method (FEM) techniques \cite{DalNegro_Crystals2019}. The validity of this approximation depends on the scatterer material and the size parameter $kR$, where $k$ is the wavelength number and $R$ is the scatterer radius \cite{DalNegro_Crystals2019}. In the following we briefly review some aspects of this formalism and we refer to Ref.\cite{DalNegro_Crystals2019} for a more detailed derivation and validation of this method.

The starting point to derive the EMCDA approximation is to evaluate the total electric and magnetic fields at the $\emph{i-th}$ particle location resulting from the electric and magnetic dipole moments of the $\emph{j-th}$ particle. These coupled equations can be written as:
\begin{equation}\label{GreenDefinition}
\begin{bmatrix}
\bm{E}_i \\
\bm{H}_i
\end{bmatrix}
=
\begin{bmatrix} 
C_{ij}	&	-f_{ij}\\
f_{ij}	&	C_{ij}\\
\end{bmatrix}
\begin{bmatrix} 
\tilde{\alpha}_E	&	0\\
0	&	\tilde{\alpha}_H\\
\end{bmatrix}
\begin{bmatrix}
\bm{E}_j \\
\bm{H}_j 
\end{bmatrix},
\end{equation}
where $\tilde{\alpha}_E$ and $\tilde{\alpha}_H$ are 3$\times$3 diagonal matrices containing the electric and magnetic polarizability $\alpha_E$ and $\alpha_H$\cite{Yurkin,Doyle}. The matrices $C_{ij}$ and $f_{ij}$ are defined by the relations 
\begin{equation}
C_{ij}=
\begin{bmatrix}
a_{ij}+b_{ij}(n_{ij}^x)^2 & b_{ij}n_{ij}^xn_{ij}^y    & b_{ij}n_{ij}^xn_{ij}^z    \\
b_{ij}n_{ij}^yn_{ij}^x    & a_{ij}+b_{ij}(n_{ij}^y)^2 & b_{ij}n_{ij}^yn_{ij}^z    \\
b_{ij}n_{ij}^zn_{ij}^x    & b_{ij}n_{ij}^zn_{ij}^y    & a_{ij}+b_{ij}(n_{ij}^z)^2
\end{bmatrix}
\label{C}
\end{equation}
\begin{equation}\label{F}
f_{ij}=
\begin{bmatrix}
0               & -d_{ij}n_{ij}^z &  d_{ij}n_{ij}^y \\
 d_{ij}n_{ij}^z &  0              & -d_{ij}n_{ij}^x \\
-d_{ij}n_{ij}^y &  d_{ij}n_{ij}^x &  0
\end{bmatrix}
\end{equation}
where $n_{ij}^\beta=\beta_{i}-\beta_{j}$  ($\beta=x,y,$ and $z$) are the components of the normal vector from the $jth$ to the $ith$ particle, while the coefficients $a_{ij}$, $b_{ij}$, and $c_{ij}$ are discussed in Refs.\cite{DalNegro_Crystals2019,Mulholland}. These matrices define the electric and magnetic dyadic Green's matrix $\overleftrightarrow{G}_{ij}$ that connects the electromagnetic field of the $i$-th-particle with the electromagnetic field of the $j$-th-particle:  
\begin{equation}\label{Gij}
\overleftrightarrow{G}_{ij}= \begin{bmatrix} 
C_{ij}	&	-f_{ij}\\
f_{ij} &	C_{ij}\\
\end{bmatrix}
=
\begin{bmatrix} 
\overleftrightarrow{G}^{ee}_{ij}	&	\overleftrightarrow{G}^{eh}_{ij}\\
\overleftrightarrow{G}^{he}_{ij}  &	\overleftrightarrow{G}^{hh}_{ij}\\
\end{bmatrix}.
\end{equation} 
The dyadic symbol $\overleftrightarrow{\{\cdots\}}$ is used because we are taking into account all the field components. $\overleftrightarrow{G}_{ij}$ is a $6\times6$ matrix. This formalism allows one to calculate the scattering, extinction, and absorption properties of an arbitrary geometry after applying a Foldy-Lax procedure \cite{DalNegro_Crystals2019} as discussed in the section \emph{``Scattering properties of prime arrays"}.

The EMCDA method provides fundamental physical information about the light transport properties of open scattering media that cannot be easily accessed via FEM or FDTD techniques. Indeed, in contrast to numerical mesh-based methods, the EMCDA extension not only allows one to obtain the frequency positions and lifetimes of all the scattering resonances, but also to fully characterize their spectral statistics and measurable scattering parameters, such as the density of optical states (DOS) \cite{Wang_Prime,SkipetrovDOS} reported in the Fig.\,2 of the manuscript. Specifically, to include the magnetic dipole contribution inside the Green's matrix spectral method, we have to isolate the traditional electric Green's matrix $G_{ij}$, defined by the relation
\begin{eqnarray}\label{GreeenElectric}
G_{ij}&&=\frac{3}{2}\left(1-\delta_{ij}\right)\frac{e^{ik_0r_{ij}}}{ik_0r_{ij}}\Biggl\{\Bigl[\bm{U}-\hat{\bm{r}}_{ij}\hat{\bm{r}}_{ij}\Bigr]-\\\nonumber
&&\Bigl(\bm{U}-3\hat{\bm{r}}_{ij}\hat{\bm{r}}_{ij}\Bigr)\nonumber	
\left[\frac{1}{(k_0r_{ij})^2}+\frac{1}{ik_0r_{ij}}\right]\Biggr\}
\end{eqnarray}
\cite{RusekPRE,LagendijkResonant,SkipetrovPRL,SgrignuoliVogel,SgrignuoliCL}, by rewriting the matrix (\ref{C}) in the compact form
\begin{equation}\label{eq28}
C_{ij}=a_{ij}U +b_{ij}\hat{r}_{ij}\hat{r}_{ij}
\end{equation}
where the term $\hat{r}_{ij}\hat{r}_{ij}$ has the explicit expression
\begin{equation*}
\begin{bmatrix}
(x_i-x_j)^2              &		(x_i-x_j)(y_i-y_j) 	& 		(x_i-x_j)(z_i-z_j) \\
(x_i-x_j)(y_i-y_j)     & 	(y_i-y_j)^2               	&		(y_i-y_j)(z_i-z_j)  \\
(x_i-x_j)(z_i-z_j)     &  	(y_i-y_j)(z_i-z_j)		&  		(z_i-z_j)^2 
\end{bmatrix}
\end{equation*}
divided by the factor $1/r^2_{ij}$.
Equation (\ref{eq28}) is then equivalent to 
\begin{eqnarray}
\begin{aligned}
C_{ij}&=(1-\delta_{ij}) \frac{e^{ik_0r_{ij}}}{r_{ij}}k_0^2 \\\nonumber
&\left[(U-\hat{r}_{ij}\hat{r}_{ij})-\left(\frac{1}{(k_0r_{ij})^2}+\frac{1}{ik_0r_{ij}}\right)(U-3\hat{r}_{ij}\hat{r}_{ij})\right]\\
&=\frac{2}{3}ik_0^3G_{ij}
\end{aligned}
\end{eqnarray}
\begin{figure*}[t!]
\centering
\includegraphics[width=\linewidth]{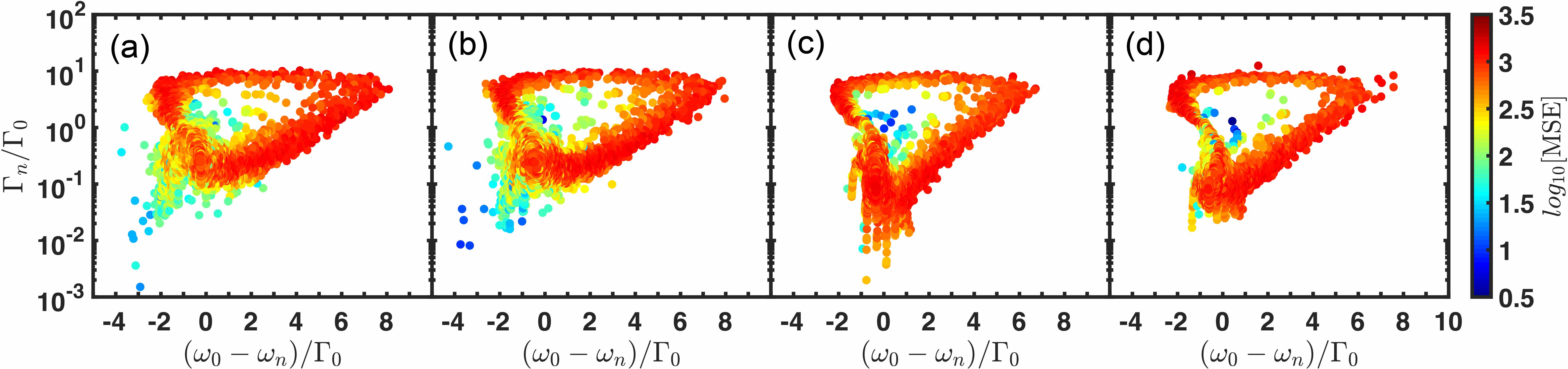}
\caption{Spectral properties of the prime arrays. Eigenvalues of the electromagnetic Green's matrix (\ref{Gfinal}) are shown by point on the complex plane for almost 2000 electric and magnetic dipoles arranged in Eisenstein (a), Gaussian (b), Hurwitz (c), and Lifschitz (d) prime geometries. All these distributions are evaluated by fixing the optical density $\rho\lambda^2$ equal to 5. This value matches with the experimental conditions discussed in the main manuscript.}
\label{FigS4}
\end{figure*}

In the same way, matrix (\ref{F}) can be rewritten in the compact form
\begin{equation}
f_{ij}=\frac{e^{ik_0r_{ij}}}{r_{ij}}k_0^2 \left(1-\frac{1}{ik_0r_{ij}}\right)\hat{R}_{ij}
\end{equation}
where 
\begin{equation*}
\hat{R}_{ij}=
\begin{bmatrix}
0             &		-\hat{r}^z_{ij} 	& 		\hat{r}^y_{ij}\\
\hat{r}^z_{ij}   & 	0               	&		-\hat{r}^x_{ij}  \\
-\hat{r}^y_{ij}     &  	\hat{r}^x_{ij}		&  0
\end{bmatrix}
\end{equation*}
denotes the cartesian components of the unit directional vector $\hat{r}_{ij}=r_{ij}/|r_{ij}|$.
Summarizing, the $6\times6$ sub-blocks of the full Green's matrix defined by equation\,(\ref{Gij}) assumes the explicit form:
\begin{eqnarray}\label{Gsemifinal}
\begin{aligned}
\overleftrightarrow{G_{ij}}&=
\begin{bmatrix} 
\overleftrightarrow{G}^{ee}_{ij}	&	\overleftrightarrow{G}^{eh}_{ij}\\
\overleftrightarrow{G}^{he}_{ij}  &	\overleftrightarrow{G}^{hh}_{ij}\\
\end{bmatrix}=\\
&=
\begin{bmatrix} 
\frac{2}{3}ik_0^3G_{ij}													&	-\Xi_{ij}\\
\Xi_{ij}             			&	\frac{2}{3}ik_0^3G_{ij}\\
\end{bmatrix}
\end{aligned}
\end{eqnarray}
where $\Xi_{ij}$ is equal to:
$$
\frac{e^{ik_0r_{ij}}}{r_{ij}}k_0^2 \left(1-\frac{1}{ik_0r_{ij}}\right)\hat{R}_{ij}
$$

To be consistent with the notation of equation\,(\ref{GreeenElectric}), we have to multiply the matrix $\overleftrightarrow{G}_{ij}$ by the factor $3/2ik_0^3$ such that:

\begin{equation}\label{Gfinal}
\frac{3}{2}\frac{1}{ik_0^3}\overleftrightarrow{G_{ij}}=
\begin{bmatrix} 
G^{ee}_{ij}	&	G^{eh}_{ij}\\
G^{he}_{ij}       &	G^{hh}_{ij}\\
\end{bmatrix}
\end{equation}
where $G^{ee}_{ij}=G^{hh}_{ij}=G_{ij}$ are defined by equation (\ref{GreeenElectric}), while the off-diagonal terms $G^{eh}_{ij}= -G^{he}_{ij}$ become 
\begin{equation}
G^{he}_{ij}=\frac{3}{2}\frac{e^{ik_0r_{ij}}}{ik_0r_{ij}}\left(1-\frac{1}{ik_0r_{ij}}\right)\hat{R}_{ij}  
\end{equation}

The matrix (\ref{Gfinal}) is a non-Hermitian matrix. As a consequence, it has complex eigenvalues $\Lambda_n$ with a physical interpretation related to the scattering frequency and decay time. Fig.\,\ref{FigS4} shows the eigenvalue distributions produced by numerically diagonalizing the matrix (\ref{Gfinal}) made from more than 2000 electric and magnetic dipoles arranged in the Eisenstein (a), Gaussian(b), Hurwitz (c), and Lifschitz (d) geometries. These complex pole distributions are color coded according to the $\log_{10}$ values of the mode spatial extent (MSE). The MSE characterizes the spatial extent of a photonic mode \cite{SgrignuoliACS}. All these data are obtained by fixing the optical density $\rho\lambda^2$ equal to 5. Here $\rho$ is the number of particles per unit area while $\lambda$ in the optical wavelength. This value is compatible with our experimental conditions discussed in the main manuscript. Moreover, the distribution of level spacing calculated from the eigenvalues of the matrix (\ref{GreeenElectric}) is reproduced by the critical cumulative probability distribution at this optical regime \cite{Wang_Prime}. This feature is confirmed by the main result of our manuscript: MF characterizes the scattering resonances of prime arrays over a broad frequency range.
\begin{figure*}[t!]
\centering
\includegraphics[width=15cm]{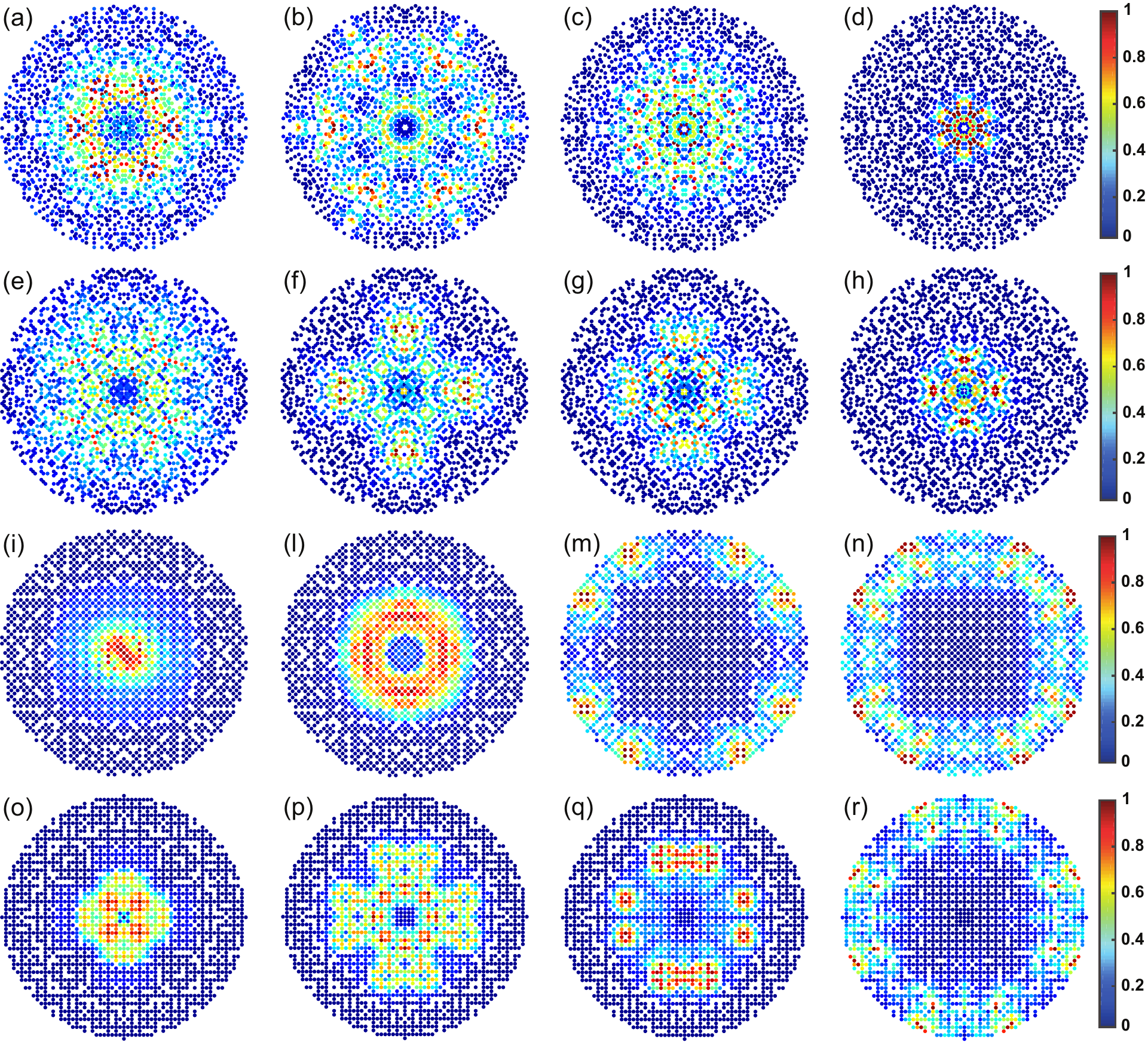}
\caption{Calculated scattering resonances of the prime arrays. Representative spatial distributions of the electromagnetic Green's matrix eigenvectors produced by approximately 2000 electric and magnetic point dipoles arranged in Eisenstein (a-d), Gaussian (e-h), Hurwitz (i-n), and Lifschitz (o-r) prime arrangement, respectively.}
\label{FigS5}
\end{figure*}

A peculiar feature is shown in Fig.\,\ref{FigS4}: the existence of spectral gaps. This characteristic, which does not occur in traditional random arrays, reflects the role played by the structural correlations of prime arrays and becomes prominent in the DOS behavior. The DOS distributions are calculated from the histograms of the real part of the eigenvalue of the matrix (\ref{Gfinal}). Within the Green's formalism, the size and the refractive index of the particles can be taken into account after the diagonalization of the Green's matrix by extracting the frequency $\omega_0$ and the decay rate $\Gamma_0$ from the central position and the lineshape of the scattering cross section of a single particle. In the present case, we have extracted these parameters from the scattering cross section of a single TiO$_2$ nanoparticle of 105\,nm radius embedded in an effective refractive index medium ($n\sim2.5$) to emulate the presence of the quartz substrate. Although the relation between the fluctuations of scattered intensity and the DOS is not trivial because it depends on many intrinsic and extrinsic factors \cite{Yasumoto}, the results of Fig.\,2 show qualitative agreement confirming the analysis provided by the frequency-frequency correlation matrix $C(\nu,\nu^\prime)$.

The spatial distribution of the scattering resonances can be evaluated by analyzing the eigenvectors of the matrix (\ref{Gfinal}). Representative scattering resonances of prime arrays are reported in Fig.\,\ref{FigS5}. These quasi-modes populate the region of the complex plane near the spectral gaps and are characterized by a low decay rates ($\Im[\Lambda_n]<1$) and a high value of the mode spatial extent ($\log_{10}$[MSE]$>1.5$). These eigenvectors are critical scattering resonances because they are long-lived and extended quasi-modes \cite{Wang_Prime}. Moreover, they show the same characteristic of the experimental dark-field images reported in the main manuscript. Specifically, the scattering resonances of Eisenstein (a-d) and Gaussian (e-h) configurations display a clear transition from quasi-modes localized at the center of the arrays to resonances more radially spread around the plane of their geometrical supports. Furthermore, band-edges scattering resonances are clearly visible in both the Hurtwitz and Lifschitz array.
\section*{Multifractal Analysis of the scattering resonances of the prime arrays}
\begin{figure*}[b!]
\centering
\includegraphics[width=16cm]{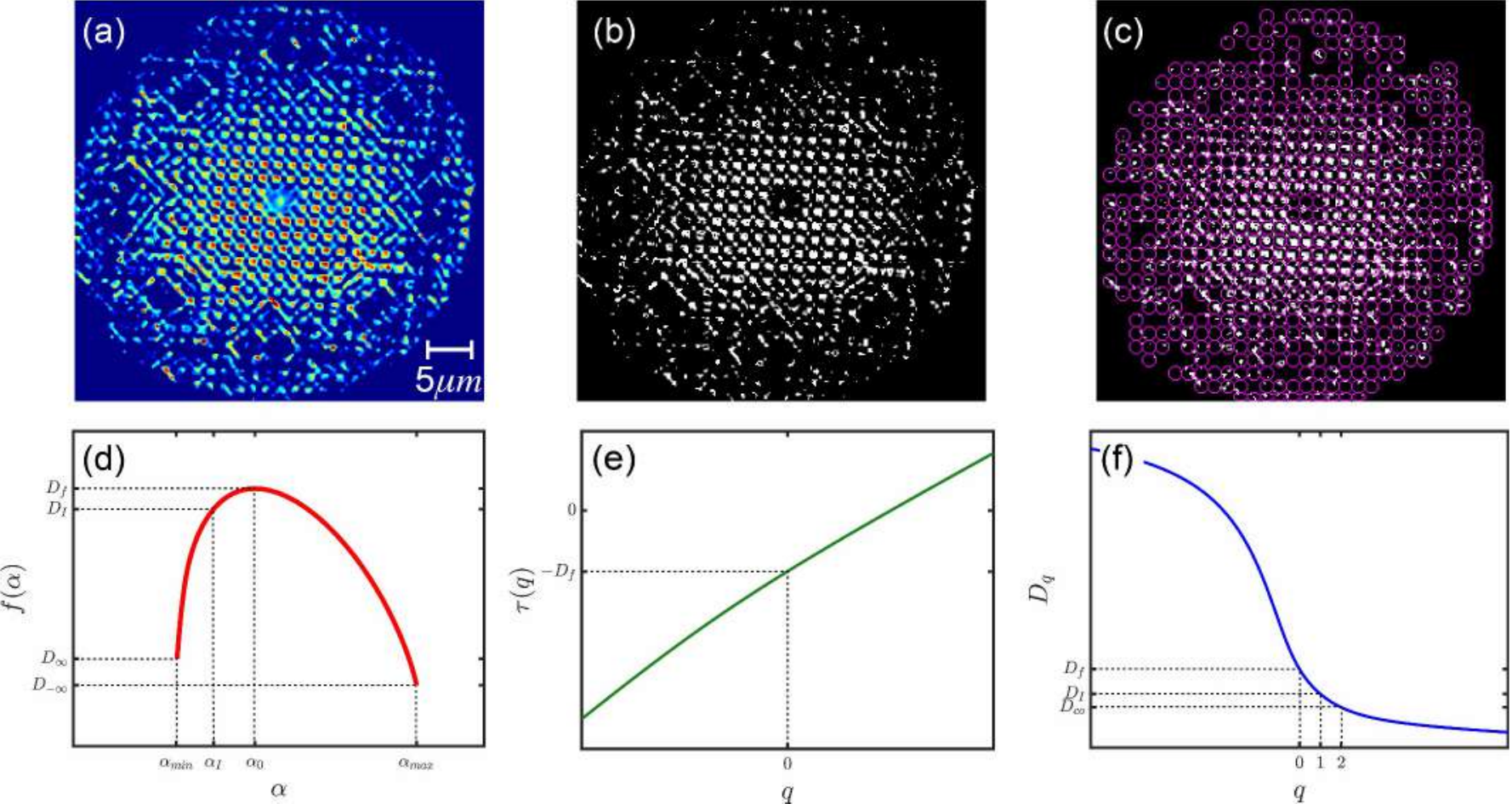}
\caption{Multifractal analysis: method overview. (a) Selected scattering resonance produced by the Hurwitz prime array at the frequency $\nu=387$THz. (b) Corresponding 600dbi bitmap image. (c) Box counting discretization. Multifractal exponents are evaluated by using the direct Chhabra-Jensen algorithm \cite{Chhabra} implemented in the FracLac routine \cite{Schneider}. Specifically, panel (d), (e), and (f) displays the multifractal singularity spectrum $f(\alpha)$, the mass exponent $\tau(q)$, and the generalized dimensions $D_q$, respectively.}
\label{FigS6}
\end{figure*}

We have used the very well know FracLac routine, developed for the Image-J software package \cite{Schneider}, to perform the multifractal analysis presented and discussed in the main manuscript. This routine is based on the method introduced in Ref.\cite{Chhabra}. In the following, we will review how evaluate the singularity spectrum $f(\alpha)$, the generalized dimensions $D_q$, and the mass exponent $\tau(q)$ from the binary map of a representative scattering resonance reported in Fig.\,\ref{FigS6}\,(a-b). These parameters are calculated by using the box-counting method and are used to discriminate if a certain distribution is non fractal, fractal, or multifractal.

As discussed in the main manuscript, the idea of the box-counting method is to divide the space embedding the object into a hyper-cubic grid of boxes of size $l$, as shown in Fig.\,\ref{FigS6}\,(c) for a representative Hurwitz scattering resonance. In inhomogeneous systems, the probability $P_i(l)$, calculated as the integral of the measure over the $i^{th}$ box, scales as $P_i(l)\sim l^{\alpha_i}$, where $\alpha_i$ are named Lipschitz-Holder exponents and quantify the strength of the singularity of a measure inside the $i^{th}$ box. Similar $\alpha_i$ values can be found at different positions within a distribution of boxes. The number of boxes $N(l)$, where $P_i$ has singularity strength $\alpha$ between $\alpha$ and $\alpha+d\alpha$, scales as $N(l)\sim l^{-f(\alpha)}$ \cite{Chhabra,Halsey,Schreiber}. The fact that the subset specified by $\alpha_i$ has its own fractal dimension $f(\alpha_i)$ explains the etymology of the word multifractal. 

Multifractal sets can also be described by a parameter called generalized dimensions $D_q$. This parameter was proposed for the first time as a different measure to describe the ``strangeness" of some phase space of dissipative dynamical systems, called attractors, which exhibit chaotic behavior \cite{Grassberger}. There are three dimensions that characterize $D_q$: the capacity dimension $D_0$, the information dimension $D_I$, and the correlation dimension $D_{co}$ \cite{MandelbrotBook2,Hentschel}. The capacity dimension is independent from $q$ and provides global information of a system, \emph{i.e.} D$_0$ coincides with the box-counting dimension $D_f$ \cite{Hilborn,Voss}. On the other hand, $D_I$ is related to the information or Shannon entropy and quantifies the degree of disorder present in a distribution \cite{MandelbrotBook2,Chhabra,Gouyet}. $D_{co}$, which is mathematically equivalent to the correlation integral \cite{Grassberger}, computes the correlation of measures contained in the intervals of size $l$. The relation between these three measures is $D_{co}\leq D_{I}\leq D_0$, where the three are equal only in the case of homogeneous fractals \cite{Korvin}. The generalized dimension is related to the scaling of the $q^{th}$ moments of a distribution and it is defined as:
\begin{equation}\label{Dq}
D_q=\frac{1}{q-1}\lim_{l\rightarrow0}\Biggl[\frac{\log \mu(q,l)}{\log l}\Biggr]
\end{equation}
where $\mu(q,l)$ is equal to
\begin{equation}\label{tau_definition}
\mu(q,l)=\sum_{i} P_i(l)^q\sim l^{-\tau(q)}
\end{equation}
Equation\,(\ref{tau_definition}) can be interpreted in the following way: if the $q^{th}$ moments of $P_i(l)$ are proportional to some power $\tau(q)$ of the box size $l$, then $\mu(q,l)$ distinguishes intertwined regions which scale in different ways according to the exponent $\tau(q)$, also named mass exponent \cite{Schreiber}. Specifically, $\tau(q)=(q-1)D_q$ shows how the moments of a given distributions scale with the box size $l$. If $\tau(q)$ is a nonlinear function of $q$, we call the distribution of measures multifractal \cite{Schreiber,Nakayama}. 

The two exponents $f(\alpha)$ and $\tau(q)$ are related to each other by means of a Legendre transformation \cite{Chhabra,Halsey,Schreiber,Nakayama}:
\begin{equation}\label{LT}
\begin{aligned}
  \tau(q)&=\alpha(q)q-f[\alpha(q)]~~~~\mbox{where}~~q=\frac{df(\alpha)}{d\alpha}\\
  f(\alpha)&=q(\alpha)-\tau[q(\alpha)]~~~~~~\mbox{where}~~\alpha=\frac{d\tau(q)}{dq}
\end{aligned}
\end{equation}
This relationship reflects a deep connection with the thermodynamic formalism of statistical mechanics where $\tau(q)$ and $q$ are conjugate thermodynamic variables to $f(\alpha)$ and $\alpha$ \cite{Chhabra}. In this context, the function $\mu(q,l)$ is formally analogous to the partition function $Z(\beta)$, while $\tau(q)$ can be interpreted as the free energy. Its Legendre transform $f(\alpha)$ is thus the analogue of the entropy, while $\alpha$ is the energy $E$. Notably, the characteristic shape of $f(\alpha)$ in the parabolic approximation (see following sections for more details) resembles the functional dependence of the entropy from $E$.

In order to evaluate these exponents, we can derive $\tau(q)$ from the log-log plot of $l$ versus $\mu(q,l)$ by using a least-fit square method. Then, $D_q$ and $f(\alpha)$ can be obtained from equations\,(\ref{Dq}) and (\ref{LT}), respectively. However, the derivation of $f(\alpha)$ through the Legendre transformation suffers from numerical inaccuracies related to the differentiation to obtain the coefficients $\alpha$ \cite{Schreiber,Nakayama,Chhabra}. In order to overcome these numerical errors, the Chhabra-Jensen method can be implemented. This method computes $f(\alpha)$ directly from the data \cite{Chhabra}. By defining the one parameter family $\hat{\mu}_i(q,l)$ through the relation:
\begin{equation}\label{mu}
\hat{\mu}_i(q,l)=\frac{P_i(l)^q}{\sum_j P_j(l)^q}
\end{equation}
the singularity strength $\alpha$ and the multifractal spectrum $f(\alpha)$ are given by:
\begin{equation}\label{a_ok}
\alpha=\lim_{l\rightarrow0}\frac{\sum_{i}\hat{\mu}_i(q,l)\log[P_i(l)]}{\log l}
\end{equation} 
\begin{equation}\label{f_ok}
f[\alpha(q)]=\lim_{l\rightarrow0}\frac{\sum_i\hat{\mu}_i(q,l)\log[\hat{\mu}_i(q,l)]}{\log l}
\end{equation}
Specifically, the numerators of equations\,(\ref{a_ok}) and (\ref{f_ok}) are evaluated for each value of $q$ for decreasing box sizes. $f[\alpha(q)]$ and $\alpha(q)$ are then extrapolated from the slopes of the plots of $\sum_i\hat{\mu}_i(q,l)\log[\hat{\mu}_i(q,l)]$ and $\sum_{i}\hat{\mu}_i(q,l)\log[P_i(l)]$ versus $\log l$, respectively.
Therefore, $\alpha$ and $f(\alpha)$ are calculated directly from the data by performing a box-counting procedure. More details on the explicit derivation of equations\,(\ref{a_ok}) and (\ref{f_ok}) can be found in Refs.\cite{Nakayama,Chhabra,Schreiber}.
 
Fig.\,\ref{FigS6}\,(d) shows a representative multifractal spectrum evaluated by using the method explained above. Fig.\,\ref{FigS6}\,(d) displays a singularity spectrum which is multifractal, \emph{i.e.} smooth downward concavities with a finite width $\Delta\alpha=\alpha_{max}-\alpha_{min}$. The mass exponent $\tau(q)$ can be also evaluated directly from Fig.\,\ref{FigS6} (c), as discussed above. Fig.\,\ref{FigS6} (e) displays the non-linear dependence of the mass exponent, demonstrating that the representative scattering resonance of panel (a) is characterized by intertwined fractal sets that scale in different ways. Notice that the moment $q=0$ corresponds to the maximum of the spectrum yielding $f(\alpha_{0})=D_f$. Finally, from the knowledge of $\tau(q)$, the generalized dimensions $D_q$ can be easily calculated by the relation $D_q=\tau(q)/(q-1)$, as reported in panel (f). 
\begin{figure*}[t!]
\centering
\includegraphics[width=\linewidth]{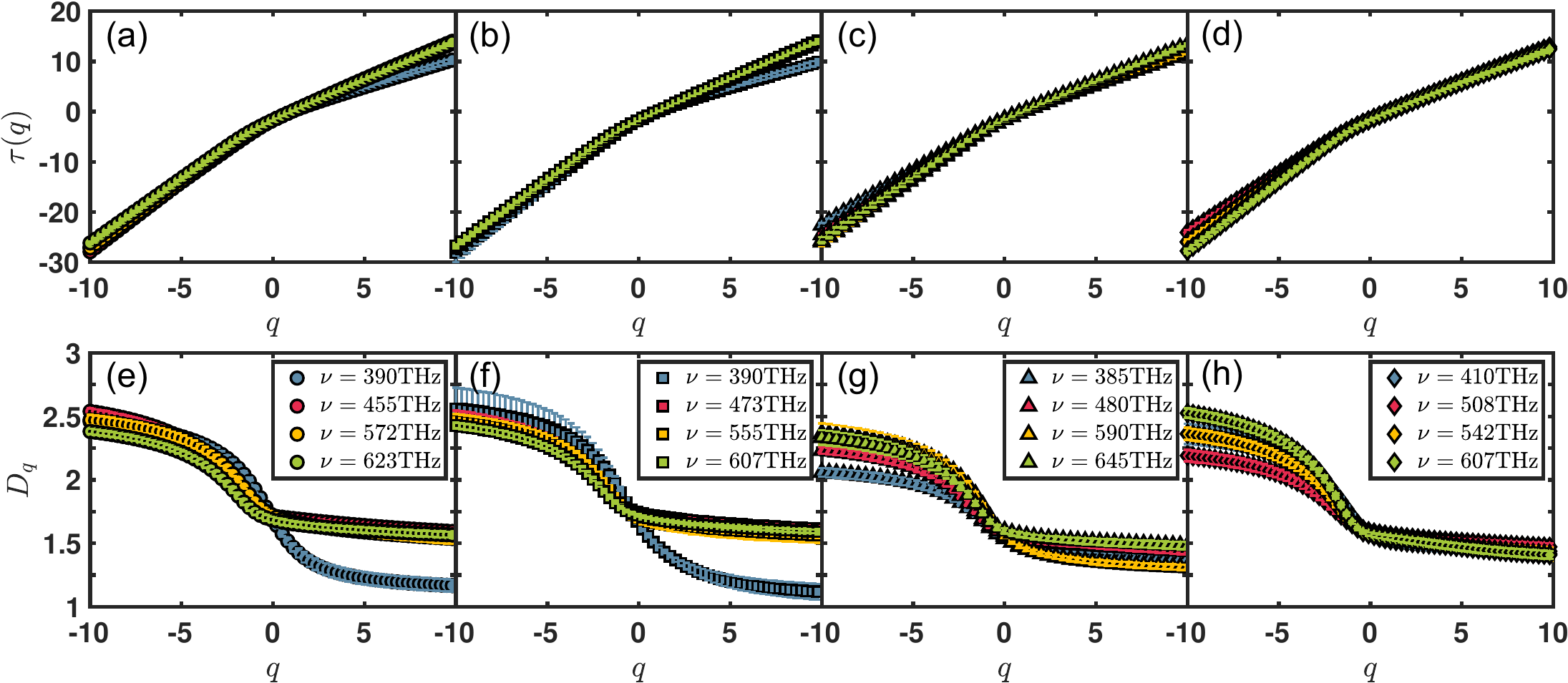}
\caption{Alternative multifractal hallmarks. Mass exponent $\tau(q)$ (a-d) and generalized dimensions $D_q$ (e-h) of the representative scattering resonances of the prime arrays discussed in Fig.\,3 of the main manuscript.}
\label{FigS7}
\end{figure*}

\section*{Mass exponent, generalized dimensions, $\Delta\alpha$, and PDF fit trends}
\begin{table*}[b!]
\caption{Generalized dimensions. Fractal ($D_f$), information ($D_I$), and correlation ($D_{co}$) dimensions of the representative scattering resonances discussed in the main manuscript.}
\centering
\def\arraystretch{2}
\begin{tabular}{|c|c|c|c|c}
\hline
      & $D_f$				  & $D_I$ 			       & $D_{co}$          &  \\
      \hline
EP &  (1.69,1.71,1.72,1.68) & (1.55,1.70,1.67,1.65) & (1.36,1.68,1.64,1.64) &  \\
\hline
GP &  (1.65,1.74,1.69,1.72) & (1.54,1.71,1.66,1.68) & (1.39,1.69,1.64,1.67) &  \\
\hline
HP &  (1.54,1.57,1.55,1.60) & (1.47,1.53,1.46,1.56) & (1.44,1.51,1.42,1.54) &   \\
\hline
LP &   (1.57,1.59,1.58,1.57) & (1.53,1.56,1.54,1.54) & (1.51,1.54,1.52,1.53) &   \\
\hline
\end{tabular}\label{Table}
\end{table*}
The exponent $\tau(q)$, as discussed in the previous section, describes the scaling of the partition function $\mu(q,l)$ with respect to box size $l$ and defines the generalized dimension through the relation $D(q)=\tau(q)/(q-1)$. Multifractals are unambiguously characterized by nonlinear $\tau(q)$ functions \cite{Stanley} and a smooth $D(q)$ function  \cite{Hentschel}. This is reported in Fig.\,\ref{FigS7} (a-d) for $\tau(q)$ and (e-h) for $D(q)$, demonstrating the multifractal nature of dark-field images of Fig.\,3 of the main manuscript. Moreover, Table \ref{Table} reports the capacity, the information, and the correlation dimensions for all the analyzed scattering resonances. These values are in agreement with the main results discussed in the manuscript. In particular, the information dimension is always larger for the Eisenstein and Gaussian arrays. This is consistent with the more inhomogeneous nature of these arrays with respect to the Hurwitz and Lifschitz arrays.

\begin{figure*}[t!]
\centering
\includegraphics[width=\linewidth]{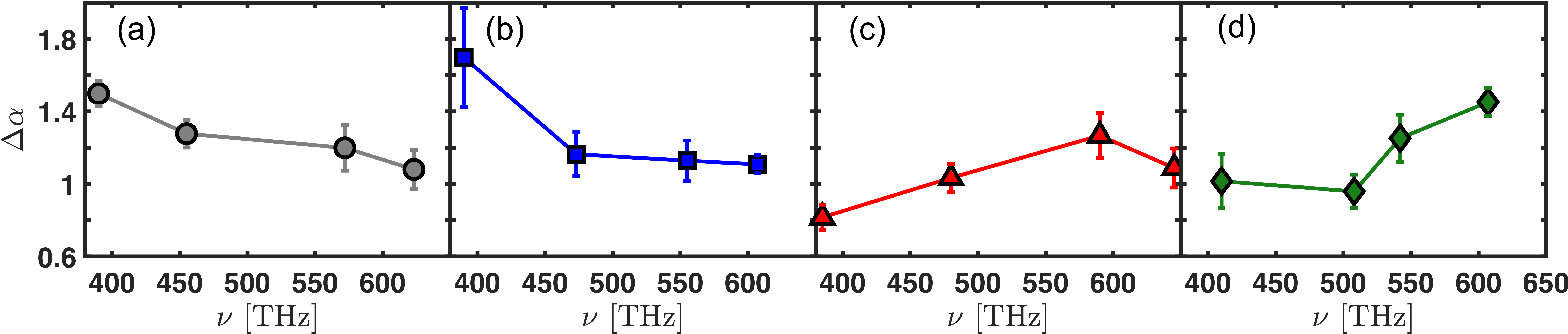}
\caption{$\Delta\alpha$ characterization. The width $\Delta\alpha$=$\alpha_{max}-\alpha_{min}$ of the multifractal spectra of Eisenstein (a), Gaussian (b), Hurwitz (c), and Lifschitz (d) arrays are reported as a function of the frequency. Here $\alpha_{min}$ (respectively $\alpha_{max}$) corresponds to the weakest (respectively the strongest)  singularity.}
\label{FigS8}
\end{figure*}

Also of interest are the $\Delta\alpha$ trends extrapolated from the multifractal spectra of Fig.\,4\,(a-d) and reported in Fig.\,\ref{FigS8}. $\Delta\alpha$ is a parameter related to the inhomogeneity of a system \cite{Richardella} and it displays, as a function of the frequency, similar features in the Eisenstein and Gaussian configuration as well as in the Hurtwitz and Liftschitz arrays resembling the classification of the prime arrays in two different classes. Indeed, while a trend from large to small $\Delta\alpha$ values is observed in the Eisenstein and Gaussian configuration, the $\Delta\alpha$ of the scattering resonances of Hurtwitz and Liftschitz arrays shows the opposite behavior.

\begin{figure*}[b!]
\centering
\includegraphics[width=\linewidth]{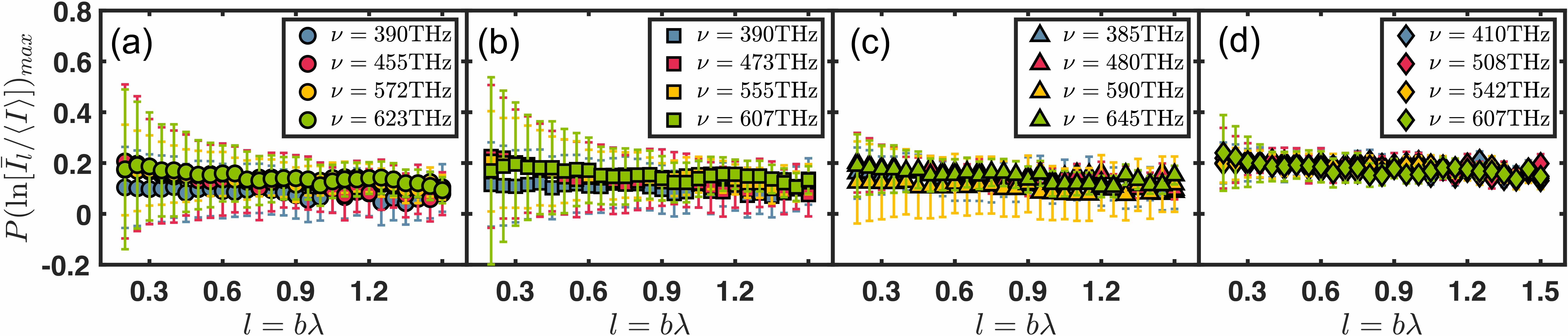}
\caption{Results of intensity PDF fits. The peak position (symbols) and the full width half maximum (bars) of the intensity histograms, normalized with respect to their averaged values, with respect to their averaged values are plotted for Eisenstein (a), Gaussian (b), Hurwitz (c), and Lifschitz (d) arrays as a function of the box size $l$ for different frequencies corresponding to the main features of their correlation matrices $C(\nu,\nu^\prime)$.}
\label{FigS9}
\end{figure*}

Finally, Fig.\,\ref{FigS9} plots the peak-position and the width of the histogram of the logarithm of the box-integrated intensity, normalized with respect to their averaged values, as a function of the box size $l=b\lambda$. These results demonstrate the stability of the non-Gaussian nature of the PDF reported in Fig.\,4\,(e-h). Moreover, the peaks of these PDFs are always shifted from their averaged intensity resembling the criticality near the Anderson transition of ultrasound waves \cite{Faez}.

\section*{Parabolic approximation of $f(\alpha)$}
Taking inspiration from the physics of the metal-insulator transition occurring in disordered electron systems \cite{Richardella,Schreiber,Evers} and from its elastic wave counterpart \cite{Faez}, we discuss in this section the consequences of the parabolic approximation applied to the scattering resonances of the prime arrays. As discussed in the main manuscript, first order perturbation theory for an Anderson transition in $2+\epsilon$ dimensions \cite{Wegner}, produces the parabolic approximation to describe the multifractal nature of its wave functions \cite{Wegner,Aoki,Castellani}. As a consequence, an infinite set of exponents, contrary to the usual situation occurring in standard critical phenomena where only two independent exponents are required \cite{StanleyBook}, have to be considered. Therefore, this critical transition is accompanied by a broadening of distributions of physical quantities governed, no longer by a unique length scale, but by a distribution of them \cite{Ioffe}. 

To understand this approximation is sufficient to remember that the multifractal analysis can be operationally implemented considering the power law behaviors of the q-th power of multiresolution structure quantities $S(q,a)\sim{a^{\zeta(q)}}$ that depend on $q$ as well as on the analysis scale $a$.
The scaling exponent $\zeta(q)$ is ten expanded to a polynomial as:
\begin{equation}\label{Expansion}
\zeta(q)=c_1q+c_2q^2/2+c_3q^3/6+\cdots
\end{equation}
The meaning of equation\,(\ref{Expansion}) is the following: (i) if $\zeta(q)$ is a linear function of $q$, then the distribution is fractal; (ii) if $\zeta(q)$ is a quadratic function of $q$ (\emph{i.e.} parabolic function of $q$), then the distribution is called weak multifractal \cite{Faez}; (iii) if $\zeta(q)$ is characterized by higher $q$ moments, then the distribution is multifractal in the strong sense. 

In order to test the deviation of an arbitrary $f(\alpha)$ with respect to the parabolic approximation: 
\begin{equation}\label{PA}
f(\alpha)=D_f-(\alpha-\alpha_0)^2/4(\alpha_0-d)
\end{equation}
we can derive a relation between the singularity spectrum and the probability distribution function of the corresponding measure \cite{Nakayama}. Denoting a distribution function of box measures $\mu_b$, grained by a box of size $l_i$, with $P(\mu_b,\delta_i)$ ($\delta_i=l_i/L$ where $L$ is the system size), $P(\mu_b,\delta_i)d\mu_b$ expresses the number of boxes that have $\mu_b$ lying in the interval $[\mu_b,\mu_b+d\mu_b]$. By defining the exponents $\alpha_i=\log\mu_b/\log\delta_i$ and if $\delta_i$ are small, we can replace $P(\mu_b,\delta_i)$ with the distribution function $\bar{P}(\alpha_i,\delta_i)$ such to define the expression $K(\log\mu_b,\delta_i)$ through the relation:
\begin{equation}\label{Dist}
K(\log\mu_b,\delta_i)=\frac{\bar{P}(\alpha_i,\delta_i)}{\log\delta_i}\approx e^{f(\alpha_i)\log\delta_i}
\end{equation}
where we have used $\bar{P}(\alpha_i,\delta_i)=\rho(\alpha_i) \delta_i^{-f(\alpha_i)}$ ($\rho(\alpha_i)$ is the density of boxes with exponent $\alpha_i$) and the fact that both $\rho(\alpha_i)$ and $1/\log \delta_i$ are weaker than $\delta_i^{f(\alpha_i)}$ for small $\delta_i$ \cite{Nakayama}.
\begin{figure*}[b!]
\centering
\includegraphics[width=\linewidth]{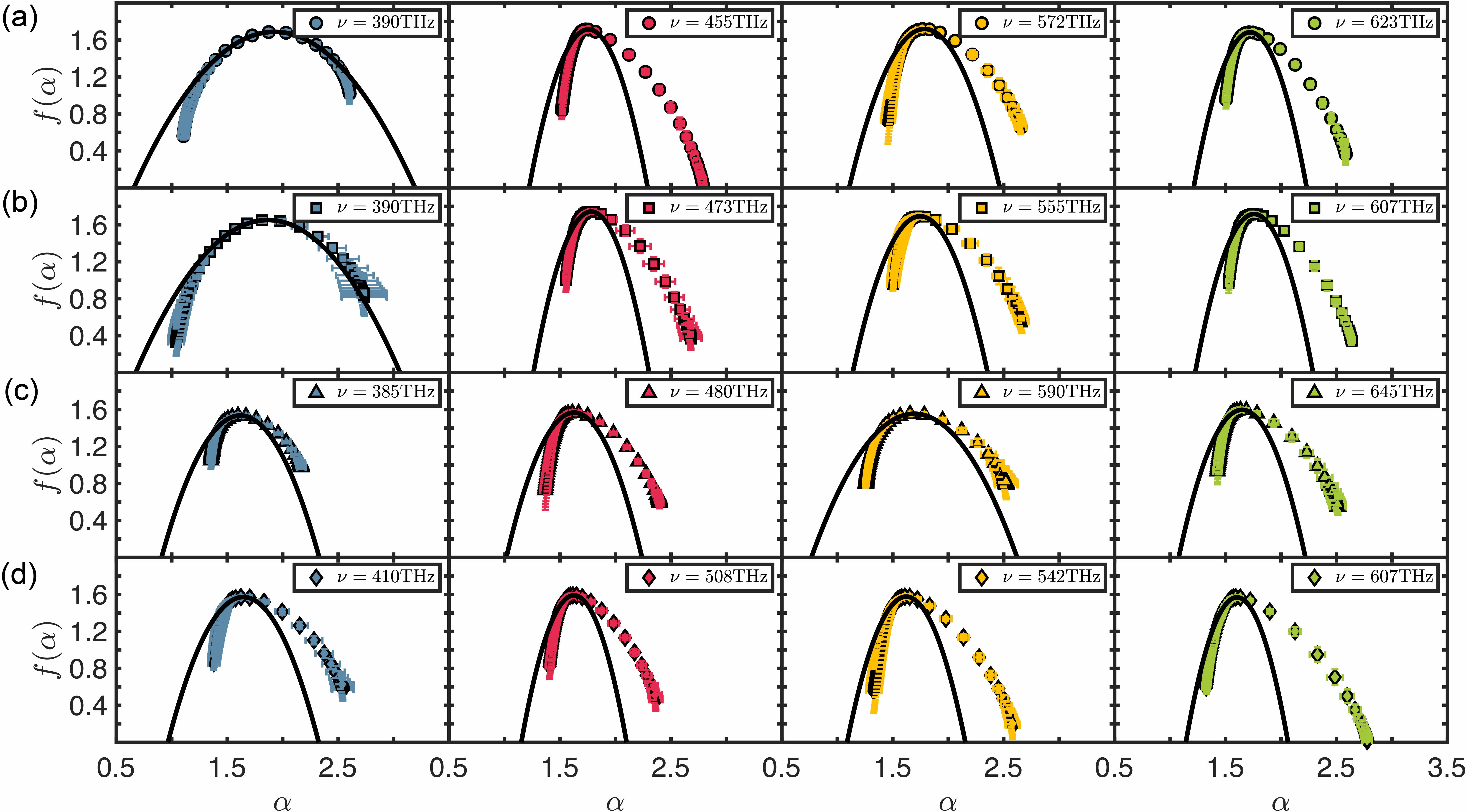}
\caption{Parabolic approximation. The parabolic approximation (black lines) is compared with respect to the multifractal spectra (symbols) evaluated directly from the experimental dark-field images of Fig.\,3 of the main paper.}
\label{FigS10}
\end{figure*}

Substituting equation\,(\ref{PA}) into equation\,(\ref{Dist}) and introducing a normalization factor $N(\delta_i)$, the distribution function $K(\log\mu_b,\delta_i)$ can be re-written as:
\begin{eqnarray}
\begin{aligned}\label{eq}
K(\log\mu_b,\delta_i)&=\frac{1}{N(\delta_i)}\exp\{-f(\alpha_i)\log\delta_i\}\\
&=\frac{1}{N(\delta_i)}\exp\left\{\left[D_f\frac{(\alpha_i-\alpha_0)^2}{4(\alpha_0-D_f)}\right]\log\delta_i\right\}\nonumber\\
\end{aligned}
\end{eqnarray}
\begin{equation}
\sim\exp\left\{\frac{(\log\mu_b-\alpha_0\log\delta_i)^2}{4(\alpha_0-D_f)\log\delta_i}\right\}
\end{equation}
This is a log-normal distribution function. Indeed, considering that $\log\delta_i<0$ and defining $x=\mu_b$, $\bar{x}=-\alpha_i|\log\delta_i|$, and $\sigma=\sqrt{2(\alpha_0-D_f)|\log\delta_i|}$, equation\,(\ref{eq}) can be rewritten in the from 
\begin{equation}\label{final}
K(\log\mu_b,\delta_i)\sim\exp\left\{-\frac{(\log x-\bar{x})^2}{2\sigma^2}\right\}
\end{equation}
Equation\,(\ref{final}) allows to evaluate the parameter $\alpha_0$, that characterizes the parabolic approximation (\ref{PA}), through the log-normal fit of the probability density function associated to the box measure $\mu_b$. 

Fig.\,\ref{FigS10} shows the deviation of the multifractal spectra $f(\alpha)$ of representative scattering resonances of prime arrays from their parabolic approximation. 
Here the box measures $\mu_b$ are the integrated intensities $I_{l_i}$. Notice that the black curves of Fig.\,\ref{FigS10} are not fits but are produced by evaluating equation\,(\ref{PA}) with the different single parameters $\alpha_0$ derived from the fit results reported in Fig.\,4\,(e-h). As clearly shown in Fig.\,\ref{FigS10}, the parabolic approximation reproduces very well the ``true" $f(\alpha)$ spectra quite well near $\alpha=\alpha_0$ but deviates significantly away from it. This implies that our data cannot be reproduced by taking into account only the quadratic term of the expansion (\ref{Expansion}). Therefore, our results demonstrate the strong multifractal character of the optical resonances of the prime arrays.
\section*{Scattering properties of prime arrays}
\begin{figure*}[t!]
\centering
\includegraphics[width=\textwidth]{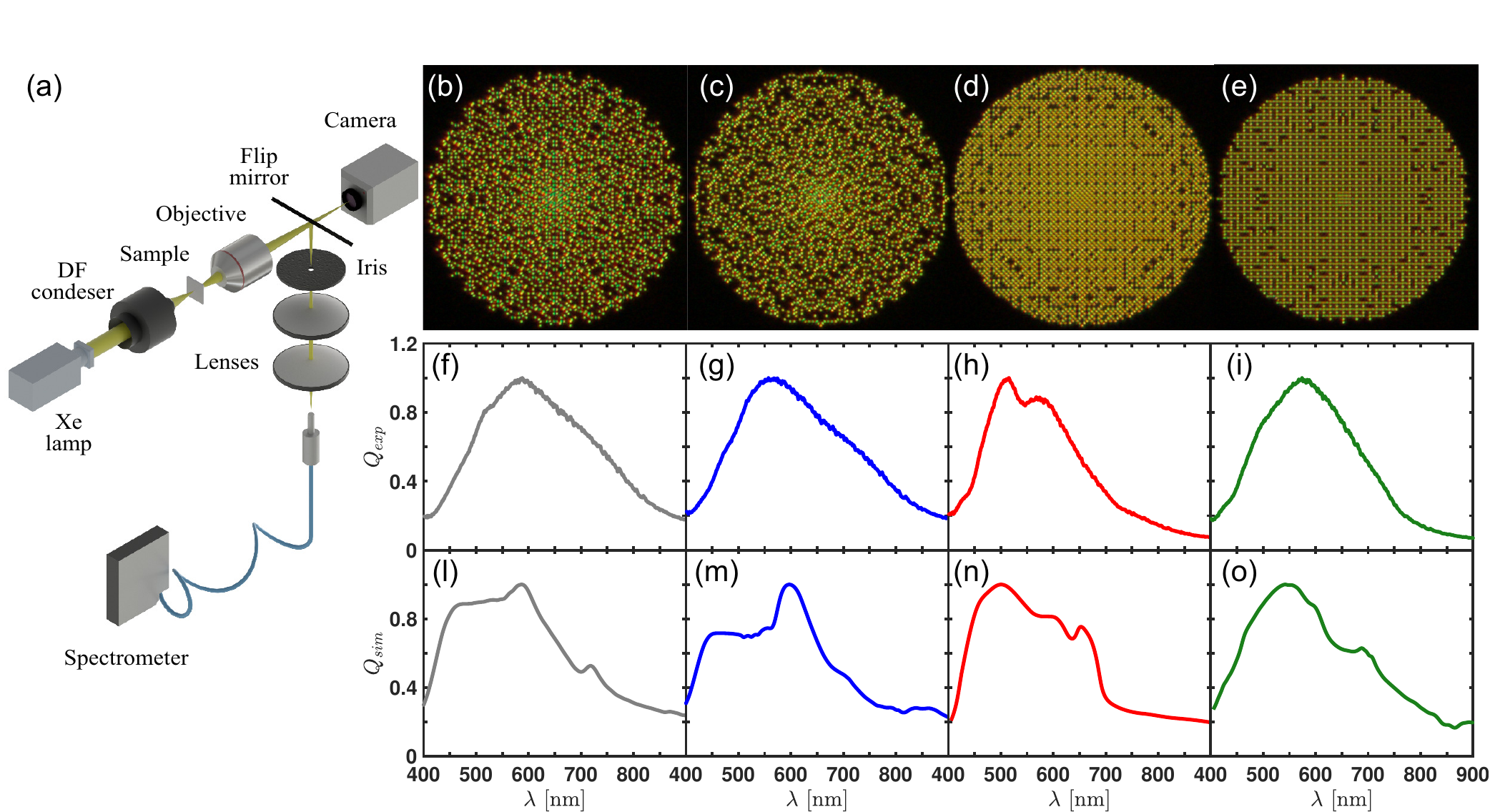}
\caption{Scattering properties of prime arrays. (a) Experimental setup to perform dark-field microscopy  measurements and dark-field scattering spectra. Multispectral dark-field  images of Eisenstein (b), Gaussian (c), Hurwitz (d), and Lifschitz (e) prime arrays. Measured (f-i) dark-field scattering spectra and calculated (l-o) scattering efficiencies of the corresponding array at the top of each column.}
\label{FigS11}
\end{figure*}

In this section we report the analysis of the scattering spectra of prime arrays. Dark-field (DF) microscopy measurements were performed using the experimental setup sketched in Fig.\,\ref{FigS11}\,(a). A 75W Xenon lamp (Olympus U-LH75XEAPO) illuminates an Olympus Darkfield Condenser with numerical aperture (NA) of 0.8-0.92 (U-DC9) which is focused onto the arrays. The incident angle cone of illumination was approximatively of $15^\circ$ with respect to the normal of the array plane. The scattered light was collect with a 0.75 NA Objective (Olympus MPLA-N) then split by a 50/50 beam splitter. On one path, the light was focused onto a CCD array to produce the dark-field images reported in Fig.\,\ref{FigS11}\,(b-e). On the other path, an image was formed onto an iris to restrict the collected light to that of a single array, and then focused into an Ocean Optics (QE65000) fiber coupled spectrometer to generate the dark-field scattering spectra shown in Fig.\,\ref{FigS11}\,(f-i). All the scattering spectra were background corrected by subtraction of the scattering signal from an equal-size unpatterned area adjacent to each prime array. In order to couple light through the microscope objective with no sample in place, the DF objective was severely defocused. Moreover, the scattering spectra were additionally corrected with respect to the normalized emission line shape of the excitation lamp. 

When the prime arrays are illuminated by a white light source, they give rise to highly organized structural color patterns as shown in Fig.\,\ref{FigS11}\,(b-e). The formation of these multispectral complex patterns, especially in the Eisenstein and Gaussian configurations, is the result of a redistribution of the incident radiation field intensity, at each given wavelength, into a multitude of spatial directions. These spatial colorimetric patterns are, therefore, the result of the complex superposition of the scattered fields that strongly depend on the nature of the optical modes generated by the interaction of light with the multifractal geometrical support of these arrays. Interestingly, these unique features were employed in the engineer of optical devices, based on Gaussian prime distributions, where both the spectral and spatial information encoded in the scattered fields were retrieved for efficient detection of protein monolayers \cite{Lee}.

Fig.\,\ref{FigS11}\,(f-i) show the measured normalized scattering intensity of prime arrays. These spectra reveal variations of the array scattering intensity as a function of the incident wavelength. Specifically, all the structures display a broad spectrum in the range (500\,nm,650\,nm) except for the Hurwitz in which two clear peaks are visible. These scattering peaks are dominated by the electric and magnetic dipole terms of the multipolar expansions \cite{Mulholland,DalNegro_Crystals2019}. Moreover, the spectral mixing of dipolar modes makes these arrays relatively insensitive to increasing the average interparticle separation. The absence of a red-shift of the scattering band as a function of interparticle separation, related to the coherent contribution of radiative coupling \cite{TrevinoNano}, is consistent with the inhomogeneity and the multifractal complexity of their geometrical supports.

To support the interpretation of our experimental results, we have evaluate the scattering efficiencies (the ratio of the scattering cross section to the volume of the particles) of prime arrays by employing the EMCDA approximation. These results, reported in Fig.\,\ref{FigS11}\,(l-o), are evaluated as the difference of the extinction and the absorption efficiency, $i.e.$ $Q_{sca}=Q_{ext}-Q_{abs}$ \cite{DalNegro_Crystals2019}. The simulated prime arrays are characterized by more than 2000 nanospheres of 105\,nm in radius separated by an average center-to-center separation of 450\,nm embedded in an effective refractive index medium to emulate the presence of a quartz substrate. These results, which approximate the nanocylinder shapes using spherical particles of 105\,nm radius, are sufficiently well-described by taking into account only the electric and magnetic dipole terms, while the contribution of higher order electromagnetic multipoles can be safely neglected for incident wavelength larger than $\sim$470\,nm. To match the experimental conditions, we have used unpolarized excitation and we have performed an average over three different angles of incidence ($\theta_{inc}$=50$^\circ$,55$^\circ$,65$^\circ$-consistent with our experimental conditions). Fig.\,\ref{FigS11}\,(l-o) reproduce qualitatively the main features of the measured scattering spectra of prime arrays. The discrepancy between simulation and experiment is attributed to the combined effects of the non-collinear nature of the dark-field excitation (limiting frequency resolution) and to the non-spherical shape of the fabricated nanocylinders.
%
\end{document}